\def\yyy{}\def\lh{\hat{l}}
\input harvmac
\input epsf
\noblackbox

\def\la{\lambda}\def\zh{\hat{z}}

\def\cW{{\cal W}}

\def\hx#1{{\hat{#1}}}

\def\Ds{\Delta^\star}\def\D{\Delta}
\def\abstract#1{
\vskip .5in\vfil\centerline
{\bf Abstract}\penalty1000
{{\smallskip\ifx\answ\bigans\leftskip 2pc \rightskip 2pc
\else\leftskip 5pc \rightskip 5pc\fi
\noindent\abstractfont \baselineskip=12pt
{#1} \smallskip}}
\penalty-1000}
\def\us#1{\underline{#1}}
\def\hth/#1#2#3#4#5#6#7{{\tt hep-th/#1#2#3#4#5#6#7}}
\def\nup#1({Nucl.\ Phys.\ $\us {B#1}$\ (}
\def\plt#1({Phys.\ Lett.\ $\us  {B#1}$\ (}
\def\cmp#1({Comm.\ Math.\ Phys.\ $\us  {#1}$\ (}
\def\prp#1({Phys.\ Rep.\ $\us  {#1}$\ (}
\def\prl#1({Phys.\ Rev.\ Lett.\ $\us  {#1}$\ (}
\def\prv#1({Phys.\ Rev.\ $\us  {#1}$\ (}
\def\mpl#1({Mod.\ Phys.\ Let.\ $\us  {A#1}$\ (}
\def\ijmp#1({Int.\ J.\ Mod.\ Phys.\ $\us{A#1}$\ (}
\def\br{\hfill\break}\def\noi{\noindent}
\def\cx#1{{\cal #1}}\def\IP{{\bf P}}\def\IZ{{\bf Z}}\def\IC{{\bf C}}
\def\tx#1{{\tilde{#1}}}\def\bb#1{{\bar{#1}}}
\def\fc#1#2{{#1 \over #2}}\def\p{\partial}
\def\frac#1#2{{#1 \over #2}}\def\p{\partial}
\def\be{\beta}\def\om{\omega}\def\Om{\Omega}
\def\subsubsec#1{\ \br \noindent {\it #1} \br}
\def\eps{\epsilon}\def\th{\theta}\def\zh{\hx z}
\def\M{\Lambda}\def\N{\Lambda^*}

\def\zk{z_{\cx H}}
\def\WP{{\bf WP}}\def\IX{{\bf X}}
\def\tablecaption#1#2{\kern.75truein\lower0truept\hbox{\vbox{\hsize=5truein\noindent{\bf Table\hskip5truept#1:} #2}}}
\def\De{\D_\flat}\def\Des{\Ds_\flat}\def\nuh{\rho}
\def\vs{\vskip5pt}

\def\-{\hphantom{-}}
\def\Wh{{W_{\cx H}}}\def\ga{\gamma}\def\lb{{\underline{l}}}

\def\Winst{\cW_{inst}}\def\vq{{q}}\def\vbe{{\be}}
\def\cl{^{(cl)}}\def\op{^{(op)}}

\noi
\def\vr{\vrule height 10pt depth 4pt}
\def\de{\delta}%
\def\mn{\the\secno.\the\subsecno}
\input xymatrix\input xyarrow
\parskip=4pt plus 15pt minus 1pt
\baselineskip=15pt plus 2pt minus 1pt
\lref\BBG{M.~Baumgartl, I.~Brunner and M.~R.~Gaberdiel,
  ``D-brane superpotentials and RG flows on the quintic,''
  JHEP {\bf 0707}, 061 (2007)
[arXiv:0704.2666 [hep-th]].}
\lref\KSii{J.~Knapp and E.~Scheidegger,
``Matrix Factorizations, Massey Products and F-Terms for Two-Parameter
Calabi-Yau Hypersurfaces,''
arXiv:0812.2429 [hep-th].}
\lref\WitCh{E.~Witten,
  ``Branes and the dynamics of {QCD},
  Nucl.\ Phys.\  B {\bf 507}, 658 (1997)
  [arXiv:hep-th/9706109].}
\lref\Katz{S.~Kachru, S.~H.~Katz, A.~E.~Lawrence and J.~McGreevy,
  ``Open string instantons and superpotentials,''
  Phys.\ Rev.\  D {\bf 62}, 026001 (2000)
  [arXiv:hep-th/9912151]; ``Mirror symmetry for open strings,''
  Phys.\ Rev.\  D {\bf 62}, 126005 (2000)
  [arXiv:hep-th/0006047].}
\lref\KT{A.~Klemm and S.~Theisen,
  ``Considerations of one modulus Calabi-Yau compactifications: Picard-Fuchs
  equations, Kahler potentials and mirror maps,''
  Nucl.\ Phys.\  B {\bf 389}, 153 (1993)
  [arXiv:hep-th/9205041].}
\lref\mibo{D.A. Cox and S. Katz, ``Mirror Symmetry and Algebraic Geometry'',
Mathematical Surveys and Monographs
Vol. 68, American Mathematical Society, 1999;\br
K. Hori et. al, ``Mirror Symmetry'', Clay Mathematics Monographs v.1., 2003}
\lref\hkty{ S.~Hosono, A.~Klemm, S.~Theisen and S.~T.~Yau,
  ``Mirror Symmetry, Mirror Map And Applications To Calabi-Yau Hypersurfaces,''
  Commun.\ Math.\ Phys.\  {\bf 167}, 301 (1995)
  [arXiv:hep-th/9308122].}
\lref\GHKK{T.~W.~Grimm, T.~W.~Ha, A.~Klemm and D.~Klevers,
  ``The D5-brane effective action and superpotential in N=1
  compactifications,''
  arXiv:0811.2996 [hep-th].}
\lref\Kont{M.~Kontsevich, ``Homological Algebra of Mirror Symmetry,'' 
Proc. Internat. Congress Math. {\bf 1} (1995) 120-139,
arXiv:alg-geom/9411018.}
\lref\PaWa{R.~Pandharipande, J.~Solomon and J.~Walcher, 
``Disk enumeration on the quintic 3-fold,'' arXiv.org:math/0610901.}
\lref\HL{R.~Harvey and B.~Lawson,
  ``Calibrated geometries,''
  Acta Math.\  {\bf 148}, 47 (1982).}
\lref\mafa{H.~Jockers and W.~Lerche,
``Matrix Factorizations, D-Branes and their Deformations,''
  Nucl.\ Phys.\ Proc.\ Suppl.\  {\bf 171}, 196 (2007)
  [arXiv:0708.0157 [hep-th]];\br
J.~Knapp,
``D-Branes in Topological String Theory,''
  arXiv:0709.2045 [hep-th].}
\lref\VafOS{ C.~Vafa,
  ``Extending mirror conjecture to Calabi-Yau with bundles,''
  arXiv:hep-th/9804131.}
\lref\comp{{\tt Puntos}, http://www.math.ucdavis.edu/~deloera/RECENT\_WORK/puntos2000;\br
{\tt Schubert}, http://math.uib.no/schubert;\br 
{\tt Instanton}, http://www.math.uiuc.edu/~katz/software.html}
\lref\rgkz{I.\ Gel'fand, M.\ Kapranov and A.\ Zelevinsky,
``Hypergeometric functions and toric varieties'', 
Funct. Anal. Appl. {\bf 23} no. 2 (1989), 12.}
\lref\mhvi{C.~Voisin, ``Hodge theory and complex algebraic geometry'',  
Cambridge Studies in Advanced Mathematics 76,77, Cambridge University Press;\br
C.A.M. Peters and J.H.M. Steenbrink, ``Mixed Hodge Structures'', A Series of Modern Surveys in Mathematics, Vol. 52, Springer, 2008.}
\lref\mhvii{P. Deligne, ``Theori\'e de Hodge I-III'',
Actes de Congr\`es international de Mathematiciens (Nice,1970),
Gauthier-Villars {\bf 1} 425 (1971);
Publ. Math. IHES {\bf 40} 5 (1971);
Publ. Math. IHES {\bf 44} 5 (1974);\br
J. Carlson, M. Green, P. Griffiths and J. Harris,
``Infinitesimal variations of 
Hodge structure (I)'', Compositio Math. {\bf 50} (1983), no. 2-3, 109.}
\lref\mhviii{P.\ Griffiths, ``
On the periods of certain rational integrals. I, II,'' 
Ann. of Math. {\bf 90} (1969) 460,496;
``A theorem concerning the differential equations satisfied
by normal functions associated to algebraic cycles'',
Am. J. Math {\bf 101} 96 (1979).}
\lref\WitCS{
  E.~Witten,
  ``Chern-Simons Gauge Theory As A String Theory,''
  Prog.\ Math.\  {\bf 133}, 637 (1995)
  [arXiv:hep-th/9207094].
}
\lref\WitHS{E.~Witten,
  ``New Issues In Manifolds Of SU(3) Holonomy,''
  Nucl.\ Phys.\  B {\bf 268}, 79 (1986).
}
\lref\Bat{ V.~V.~Batyrev,
  ``Dual polyhedra and mirror symmetry for Calabi-Yau hypersurfaces in toric
  varieties,''
  J.\ Alg.\ Geom.\  {\bf 3}, 493 (1994). 
}
\lref\wip{Work in progress}
\lref\AV{
  M.~Aganagic and C.~Vafa,
  ``Mirror symmetry, D-branes and counting holomorphic discs,''
  arXiv:hep-th/0012041.
}
\lref\AVp{
  M.~Aganagic and C.~Vafa, private communication.
}
\lref\glsm{
  E.~Witten,
  ``Phases of N = 2 theories in two dimensions,''
  Nucl.\ Phys.\  B {\bf 403}, 159 (1993)
  [arXiv:hep-th/9301042];
}
\lref\glsmii{
   P.~S.~Aspinwall, B.~R.~Greene and D.~R.~Morrison,
  ``Calabi-Yau moduli space, mirror manifolds and spacetime topology  change in
  string theory,''
  Nucl.\ Phys.\  B {\bf 416}, 414 (1994)
  [arXiv:hep-th/9309097].
}
\lref\HV{
  K.~Hori and C.~Vafa,
  ``Mirror symmetry,''
  arXiv:hep-th/0002222.
}
\lref\Wa{
  J.~Walcher,
  ``Opening mirror symmetry on the quintic,''
  Commun.\ Math.\ Phys.\  {\bf 276}, 671 (2007)
  [arXiv:hep-th/0605162].
}
\lref\Jo{H.~Jockers and M.~Soroush,
  ``Effective superpotentials for compact D5-brane Calabi-Yau geometries,''
  arXiv:0808.0761 [hep-th].
}
\lref\LM{ W.~Lerche and P.~Mayr,
  ``On N = 1 mirror symmetry for open type II strings,''
  arXiv:hep-th/0111113.}
\lref\LMW{ W.~Lerche, P.~Mayr and N.~Warner,
  ``N = 1 special geometry, mixed Hodge variations and toric geometry,''
  arXiv:hep-th/0208039; 
  ``Holomorphic N = 1 special geometry of open-closed type II strings,''
  arXiv:hep-th/0207259.}
\lref\PM{P.~Mayr,
  ``N = 1 mirror symmetry and open/closed string duality,''
  Adv.\ Theor.\ Math.\ Phys.\  {\bf 5}, 213 (2002)
  [arXiv:hep-th/0108229].}
\lref\OV{
  H.~Ooguri and C.~Vafa,
  ``Knot invariants and topological strings,''
  Nucl.\ Phys.\  B {\bf 577}, 419 (2000)
  [arXiv:hep-th/9912123].
}
\lref\AKV{M.~Aganagic, A.~Klemm and C.~Vafa,
  ``Disk instantons, mirror symmetry and the duality web,''
  Z.\ Naturforsch.\  A {\bf 57}, 1 (2002)
  [arXiv:hep-th/0105045].
}
\lref\Heho{
  M.~Herbst, K.~Hori and D.~Page,
  ``Phases Of N=2 Theories In 1+1 Dimensions With Boundary,''
  arXiv:0803.2045 [hep-th].
}
\lref\Forbes{  B.~Forbes,
  ``Open string mirror maps from Picard-Fuchs equations on relative
  cohomology,''
  arXiv:hep-th/0307167.
}

\Title{\vbox{
\rightline{\vbox{\baselineskip12pt
\hbox{LMU-ASC 02/09\hskip 1cm }
}}}}
{Mirror Symmetry for Toric Branes} 
\vskip-1cm\centerline{{\titlefont on Compact Hypersurfaces}}\vskip 0.3cm
\centerline{M. Alim, M. Hecht, P. Mayr and A. Mertens}
\vskip 0.6cm
\centerline{\it Arnold Sommerfeld Center for Theoretical Physics}
\centerline{\it LMU, Theresienstr. 37, D-80333 Munich, Germany}
\vskip 0cm
\abstract{\noi
{}We use toric geometry to study open string mirror symmetry on 
compact Calabi--Yau manifolds. 
For a mirror pair of toric branes on a 
mirror pair of toric hypersurfaces we derive a canonical 
hypergeometric system of differential equations, whose solutions 
determine the open/closed string mirror maps and the 
partition functions for spheres and discs. We define a 
linear sigma model for the brane geometry and describe a
correspondence between dual toric polyhedra 
and toric brane geometries.
The method is applied to study examples with obstructed and classically 
unobstructed brane moduli at various points in the deformation space. 
Computing the instanton expansion at large volume 
in the flat coordinates on the open/closed deformation space we obtain 
predictions for enumerative invariants. 
} 
\Date{\vbox{\hbox{\sl {\hskip0.8cm January 2009}}
\vskip2cm
\noi {\ninerm 
\hbox{\hskip1.2cm murad.alim, michael.hecht, mayr, adrian.mertens  at physik.uni-muenchen.de}}
}}
\goodbreak

\newsec{Introduction}
Mirror symmetry has been the subject of intense research over many years 
and its study remains rewarding. Whereas the early works focused on 
the closed string sector and the Calabi--Yau (CY) geometry, 
the interest has shifted to the interpretation of 
mirror symmetry as a duality of 
D-brane categories and the associated open string sector \Kont. 
One object of particular interest is the disc partition 
function $\cx F^{0,1}$ for an $A$ brane on a compact CY 3-fold, which depends
on the K\"ahler type deformations of the brane geometry and 
is an important datum for the definition of the category of $A$ branes. 
If a modulus is classically unobstructed, the large volume 
expansion of the disc partition function captures an
interesting enumerative problem of ``counting'' holomorphic discs
that end on the $A$ brane. In a certain parametrization motivated
by physics, the coefficients of this instanton expansion in the $A$ model
are predicted to be the {\it integral} Ooguri-Vafa invariants \OV.

One of the virtues of mirror symmetry, first demonstrated for the sphere 
partition function in \ref\can{P.~Candelas, X.~C.~De La Ossa, P.~S.~Green and L.~Parkes,
  ``A pair of Calabi-Yau manifolds as an exactly soluble superconformal
  theory,''
  Nucl.\ Phys.\  B {\bf 359}, 21 (1991)} and for the disc 
partition functions in \AV, is the ability to
compute the instanton expansion of the $A$ 
model partition function in the mirror $B$ model.
The disc partition function relates on the $B$ model side to the
holomorphic Chern-Simons functional on the CY $Z^*$  \WitCS
\eqn\csf{
S(Z^*,A)=\int_{Z^*} \rm tr (\fc{1}{2} A\wedge dA +\fc{1}{3}A\wedge A\wedge A)\wedge \Omega\ .
}
In the physical string theory $S$ represents a space-time superpotential 
obstructing some of the moduli of the brane geometry and the 
instanton expansion of the $A$ model is, under certain conditions, 
the non-perturbative superpotential generated by space-time instantons \refs{\WitHS,\OV}. While the action of mirror symmetry on the moduli space 
and the computation of superpotentials is well understood
for non-compact brane geometries\foot{See e.g. \ref\revi{M. Mari\~no, ``Chern-Simons Theory, Matrix Models, and Topological Strings'', International Series of Monographs on Physics, Oxford University Press, 2005;
\br W.~Lerche,``Special geometry and mirror symmetry for open string backgrounds with N  =
  1 supersymmetry,''
  arXiv:hep-th/0312326.} for a summary.}, the physically interesting case of 
branes on compact CY 3-folds has been elusive. 
Starting with \Wa, 
superpotentials for a class of 
involution branes without open string moduli have been studied%
\nref\MW{
  D.~R.~Morrison and J.~Walcher,
  ``D-branes and Normal Functions,''
  arXiv:0709.4028 [hep-th].}
\nref\Waii{J.~Walcher,
  ``Evidence for Tadpole Cancellation in the Topological String,''
  arXiv:0712.2775 [hep-th].}
\nref\KW{ D.~Krefl and J.~Walcher,
  ``Real Mirror Symmetry for One-parameter Hypersurfaces,''
  JHEP {\bf 0809}, 031 (2008)
  [arXiv:0805.0792 [hep-th]].}
\nref\KSi{J.~Knapp and E.~Scheidegger,
  ``Towards Open String Mirror Symmetry for One-Parameter Calabi-Yau
  Hypersurfaces,''
  arXiv:0805.1013 [hep-th].}%
\hskip-6pt in \refs{\MW-\Jo}. The 
definition of the Lagrangian $A$ brane geometry as the fixed point of
an involution has various limitations: It allows to study only
discrete brane moduli compatible with the involution and
the instanton invariants computed by the superpotential
are not generic disc invariants, but rather the number of real rational curves 
fixed by the involution \PaWa.

The present lack of a 
systematic description of the geometric deformation space 
in the compact case is a serious obstacle to the general 
study of open string mirror symmetry on compact manifolds, 
in particular the computation
of superpotentials and mirror maps for more general deformations including
open string moduli.
For the closed string case without branes, a powerful approach to study mirror symmetry
is given in terms of 
gauged linear sigma models and toric geometry \refs{\glsm,\glsmii},
in particular if combined with Batyrev's construction of dual manifolds via
toric polyhedra \Bat.\foot{We refer to \mibo\ for background material and references.} 
A similar description of open string mirror symmetry has been 
given for non-compact branes in \refs{\LM,\LMW}, 
starting from the definition of toric
branes of ref.\refs{\AV}. 
A first important step to generalize these concepts to the compact case
has been made in \Jo\ by applying the 
$\cx N=1$ special geometry defined in \LMW\ to involution branes.

The class of toric branes defined in \AV\ (see also \HL) is much larger 
then the class of involution branes and allows for relatively generic 
deformations.
The purpose of this note is to describe a  
toric geometry approach to open string
mirror symmetry for 
toric branes on compact manifolds. Specifically we consider 
mirror pairs $(Z,L)$ and $(Z^*,E)$, where $Z$ and $Z^*$ is a mirror pair of 
compact CY 3-folds described as hypersurfaces in toric varieties, and $L$ and $E$ is a
mirror pair of branes on these manifolds with a simple toric description.\foot{In the
following, $L$ will denote the $A$ brane wrapped on a Lagrangian submanifold 
and $E$ the holomorphic bundle corresponding to a $B$ brane.}
For these toric brane geometries we derive in sect.~2 a canonical 
system of differential equations
that determines the open/closed string mirror maps and the partition 
functions for spheres and discs at any point in the moduli space. 
The $B$ model geometry for this Picard-Fuchs system relates to 
a certain gauged linear sigma model, which may be associated with an ``enhanced'' 
toric polyhedron $\De$.
A dual pair of enhanced polyhedra $(\De,\Des)$ 
encodes the mirror pair of compact CY manifolds $(Z,Z^*)$ 
{\it and} the mirror pair $(L,E)$ of $A$ and $B$ branes on it, 
extending in some sense Batyrev's \Bat\ correspondence between toric polyhedra 
and CY manifolds to the open string sector.
In sect.~3 we apply this method to study 
some compact toric brane geometries with 
obstructed and classically  unobstructed moduli. The phase 
structure of the linear sigma model can be used to define 
and study large volume phases of the brane geometry, where the
superpotential has an instanton expansion in the classically
unobstructed moduli. We compute the mirror maps and the 
superpotentials and find agreement with the integrality 
predictions of \refs{\OV,\Wa} for both closed and open string deformations. 
A more complete treatment and derivations 
of some of the formula presented below are deferred to an upcoming paper \wip.

\newsec{Toric brane geometries and differential equations}
\subsec{Toric hypersurfaces and branes}
\vskip-7pt
Our starting point will be a mirror 
pair of compact CY 3-folds $(Z,Z^*)$ defined as hypersurfaces
in toric varieties $(W,W^*)$. By the correspondence of ref.\Bat, one may associate to the pair
of manifolds $(Z,Z^*)$ a pair of integral polyhedra $(\D,\D^*)$ in a four-dimensional
integral lattice $\M_4$ and its dual $\N_4$. The $k$ integral points $\nu_i(\D)$ of the polyhedron 
$\D$ correspond to homogeneous coordinates $x_i$ on the toric ambient space $W$ and 
satisfy $M=h^{1,1}(Z)$ linear relations\foot{For 
simplicity we neglect points on faces of codimension
one of $\D$ and assume that $h^{1,1}(W)=h^{1,1}(Z)$.}  
$$
\sum_{i} l^a_i \nu_i=0,\quad a=1,...,M\, .
$$
The integral entries of the vectors $l^a$ for fixed $a$ define 
the weights $l_i^a$ of the coordinates $x_i$ 
under the $\IC^*$ action
$$
x_i \to (\la_a)^{l^a_i} x_i,\qquad \la_a\in \IC^*\, , 
$$
generalizing the idea of a weighted projective space. Equivalently, the $l^a_i$ are the 
$U(1)_a$ charges of the fields in the gauged linear sigma model (GLSM)
associated with the toric variety \refs{\glsm}. The toric variety $W$  is defined as $\IC^k$ 
divided by the $(\IC^*)^M$ action and 
deleting a certain exceptional subset $\Xi$ of degenerate
orbits, $W\simeq(\IC^k - \Xi)/(\IC^*)^M$.  

In the context of CY hypersurfaces, $W$ will be the total space of the 
anti-canonical bundle over a toric variety with positive first Chern class. 
The compact manifold $Z\subset W$ is defined by
introducing a superpotential $W_Z=x_0\, p(x_i)$ in the GLSM, 
where $x_0$ is the coordinate on the fiber and $p(x_i)$ a polynomial
in the $x_{i>0}$ of degrees $-l_0^a$. 
At large K\"ahler
volumes, the critical locus is at $x_0=p(x_i)=0$ and defines the compact CY as 
the hypersurface $Z:\ p(x_i)=0$ \refs{\glsm}. 
To be concrete, we will later study $A$ branes on 
the following examples of CY hypersurfaces:
\eqn\defmfs{\vbox{\offinterlineskip\halign{
\strut # &\hfil~$#$~\hfil
&\hfil~$#$:\ ~\hfil
&~$#$~\hfil\cr
\ && 
\IX^{(1,1,1,1,1)}_5& 
\matrix{&x_0&x_1&x_2&x_3&x_4&x_5\cr (l^1)=&(-5&1&1&1&1&1\, )}\cr\cr
&& 
\IX^{(1,1,1,6,9)}_{18}& 
\matrix{&x_0&x_1&x_2&x_3&x_4&x_5&x_6\cr
(l^1)=&(-6&3&2&1&0&0&0\, )\cr
(l^2)=&(\hskip6pt 0&0&0&-3&1&1&1\, )}\cr\cr
&& 
\IX^{(1,1,1,3,3)}_9& 
\matrix{&x_0&x_1&x_2&x_3&x_4&x_5&x_6\cr 
(l^1)=&(-3&1&1&1&0&0&0\, )\cr
(l^2)=&(\hskip6pt 0&0&0&-3&1&1&1\, )}\cr
}}
}
As indicated by the notation, this is the familiar 
quintic in projective space $\IP^4=\WP^4_{1,1,1,1,1}$ in the first 
case and a degree 18 (9) hypersurface 
in a blow up of a weighted projective space $\WP^4_{1,1,1,6,9}$
($\WP^4_{1,1,1,3,3}$) in the other two cases.%
\foot{The deleted set is $\Xi=\{x_i=0, \forall i>0\}$ for 
$\IP^4$ and $\Xi=\{\{x_1=x_2=x_3=0\}\cup\{x_4=x_5=x_6=0\}\}$ in the
other two cases. The toric polyhedra will be given in sect.~3.} 

On these toric manifolds we consider a certain class of mirror pairs 
of branes, defined in \AV\ by another 
set of $N$ charge vectors $\lh^a$ for the fields $x_i$.\foot{A hat will be sometimes used to distinguish objects from the
open string sector.}
The Lagrangian submanifold wrapped by the $A$ brane $L$
is described in terms
of the vectors $\lh^a$ by the equations
\eqn\laga{
\sum_{i} \lh_i^a |x_i|^2=c_a,\qquad \sum_{i} v^i_b \th^i=0,\qquad
\sum_{i} \lh_i^a v^i_b=0\ ,
}
where $a,b=M+1,...,M+N$.
The $N$ real constants $c_a$ parametrize the brane position and the integral
vectors $v^i_b$ may be defined as a linearly independent basis of solutions
to the last equation.
As in \AV\ we restrict to special Lagrangians which requires that 
the extra charges add up to zero as well, $\sum_i \lh_i^a = 0$.

Applying mirror symmetry as in \refs{\HV,\Bat}, the mirror 
manifold $Z^*$ is defined in the toric variety $W^*$ 
by the equations 
\eqn\mmf{
p(Z^*)=\sum_{i} y_i,\qquad \prod_i y_i^{l_i^a}=z_a\ , \quad a=1,...,M.
}
The parameters 
$z_a$ are the complex moduli of the hypersurface $Z^*$ and classically
related to the complexified K\"ahler moduli $t_a$ of $Z$ by $z_a=e^{2\pi i t_a}$.
The precise relation $z_a=z_a(t_b)$ is called the mirror map and 
is generically complicated. In the open string sector, the mirror transformation of \HV\ maps the $A$ brane \laga\  to a $B$ brane $E$ 
defined by the holomorphic equations \AV%
\eqn\lagb{\cx B_a(E):\ \prod_i y_i^{\lh_i^a}-\zh_a=0,\qquad 
\zh_a=\eps_{a} e^{-c_{a}},\ a=M+1,...,M+N.}  
The (possibly obstructed) complex open string 
moduli $\zh_a$ arise from the combination of the 
phases $\eps_a$ dual to the gauge field background on the $A$ brane 
and the parameters $c_a$ in \laga \VafOS.

The class of toric branes defined above is quite general and describes
many interesting cases,
in particular involution branes with an obstructed modulus as well
as branes with classically unobstructed moduli. It is instructive
to consider the quintic $\IX^{(1,1,1,1,1)}_5$, which will be one of the
manifolds studied in sect.~3. The manifold $Z$ for the $A$ model 
is defined by a generic degree 5 polynomial in $\IP^4$, while 
the mirror manifold $Z^*$ is given in terms of eq.\mmf\ 
by the superpotential and relation\foot{The coefficients $a_i$ are 
homogeneous coordinates on the space of complex structure and related 
to the $z_a$ in \mmf\ by an rescaling of the variables $y_i$.}
\eqn\pq{
p(Z^*)=\sum_{i=0}^5 a_iy_i=0,\qquad y_1y_2y_3y_4y_5=y_0^5.
}
A change of coordinates  $y_i=x_i^5,\ i=1,...,5$ and a rescaling 
leads to the more familiar form in $\IP^4$
\eqn\qb{
p(Z^*)=\sum_{i=1}^5 x_i^5-\psi\, x_1x_2x_3x_4x_5=0,\qquad
\psi^{-5}=-\fc{a_1a_2a_3a_4a_5}{a_0^5}\equiv z_1\ .
}

The above definition of toric branes has an interesting overlap 
with more recent studies of $B$ branes via matrix factorizations\foot{See 
refs.\mafa\ for a summary.}. 
Consider the charge vectors
\eqn\cvex{\vbox{\offinterlineskip\halign{
\strut # height 12pt 
&\hfil~$#$~\hfil&
&\hfil~$#$~&\hfil~$#$~&\hfil~$#$~
&\hfil~$#$~&\hfil~$#$~&\hfil~$#$~
\vrule\cr
\noalign{\hrule}
\vrule&&x_0&x_1&x_2&x_3&x_4&x_5\cr 
\noalign{\hrule}
\vrule&l^1=&-5&1&1&1&1&1\cr
\vrule&\lh^2=&0&1&-1&0&0&0\cr
\vrule&\lh^3=&0&0&0&1&-1&0\cr
\noalign{\hrule}
}}
}
For the special values $c_a = 0$ 
the equation \laga\ for the Lagrangian submanifold can be 
rewritten as
$$
x_1=\bb x_2,\ x_3=\bb x_4,\ x_5=\bb x_5 \, .
$$
The
above equation describes an involution brane on the quintic defined as the
fixed set of the $\IZ_2$ action $(x_1,x_2,x_3,x_4,x_5)
\to(\bb x_2,\bb x_1,\bb x_4,\bb x_3,\bb x_5)$.  
The equation for the mirror $B$-brane follows from \lagb:
\eqn\exqb{
y_1=\zh_2\, y_2,\quad y_3=\zh_3\, y_4,\qquad {\rm or}\qquad x_1^5=\zh_2\, x_2^5,\quad 
x_3^5=\zh_3\, x_4^5\, .
}
A naive match of the moduli of the $A$ and $B$ model 
together with a choice of phase leads to $\zh_2=\zh_3=-1$ and the above equations become
\eqn\bbq{
x_1^5+x_2^5=0,\ x_3^5+x_4^5=0, \ (x_5^2-\psi^{1/2} x_1x_3)\, (x_5^2+\psi^{1/2} x_1x_3)\, x_5 =0  \ .
}
These equations define a set of holomorphic 2-cycles in $Z^*$ which may be 
wrapped by the 
D5 brane mirror to the $A$ brane on the Lagrangian subset defined by \laga.

Eq. \bbq\ should be compared to the results of refs.~\refs{\Wa,\MW}, 
where the 2-cycle wrapped by the $B$ brane mirror to an involution brane has 
been determined in a much more involved way along the lines of \Heho, 
by proposing a matrix factorization and computing the second algebraic 
Chern class of the associated complex. The result agrees with the 
above result from a simple application of mirror symmetry for toric branes.
A conclusive match of the toric brane defined by \lagb\
and the matrix factorization brane studied in \refs{\Wa,\MW}\ 
will given in sect.~3, where we compute the superpotential from the 
toric family and find agreement near a specific critical locus.

There are ambiguities in the above 
match between the $A$ and the $B$ model that need to be resolved
by a careful study of boundary conditions. E.g. 
in \bbq, the last equation is the superpotential intersected 
with the two hypersurfaces \exqb, but one may permute the meaning of 
the three equations. The parametrization 
\eqn\qex{\vbox{\offinterlineskip\halign{
\strut # height 12pt 
&\hfil~$#$~\hfil&
&\hfil~$#$~&\hfil~$#$~&\hfil~$#$~
&\hfil~$#$~&\hfil~$#$~&\hfil~$#$~
\vrule\cr
\noalign{\hrule}
\vrule&&x_0&x_1&x_2&x_3&x_4&x_5\cr 
\noalign{\hrule}
\vrule&l^1=&-5&1&1&1&1&1\cr
\vrule&\lh^2=&1&0&0&0&0&-1\cr
\vrule&\lh^3=&0&1&-1&0&0&0\cr
\noalign{\hrule}
}}
}
leads to the same equations \bbq\ for the special values $\hx z_2=\psi$, $\hx z_3=-1$
of the (new) moduli. An important aspect in resolving these ambiguities is
provided by the mirror map $z_a(t_b)$ on the open/closed moduli space, as it determines where a specific (family of) point(s) in the 
$A$ moduli attaches in the moduli space of the $B$ model and vice versa.
In the above example we have simply 
used the classical version of the open string mirror maps $|\hx z_a|=e^{-c_a}$ 
to find agreement with the result from matrix factorizations.
More seriously we will compute the exact mirror map --
which may in principle deviate substantially from the classical expression --
to determine the $B$ brane configuration. Since some of the 
deformations will be fixed at the critical points of the superpotential
it is in fact more natural to start with the computation
of the $B$ model superpotentials and to find its critical points. Computing 
the mirror map near these points determines a correlated set of points 
in the $A$ model parameter space,
which may or may not allow for a nice classical $A$ brane interpretation.
\vskip1cm

\subsec{$\cx N=1$ special geometry of the open/closed deformation space}
\def\gm{\nabla}\def\Si{\Sigma}
We proceed by discussing a general structure of the open/closed deformation space
that will be central to the following approach to mirror symmetry for 
the toric branes defined above.
In \refs{\LMW,\Jo} it was shown, that the open/closed string 
deformation space for 
$B$-type D5-branes wrapping 2-cycles $C$ in $Z^*$ can be studied 
from the variation of mixed Hodge structure on a deformation family of relative cohomology 
groups $H^3(Z^*,\cx H)$ of $Z^*$, where $\cx H$
is a subset that captures the deformations of $C$.\foot{Physically, 
$\cx H$ may be interpreted as a D7-brane which 
contains the D5 brane world-volume \wip.} 
In the simplest case, $\cx H$ is a single hypersurface and 
the action of the closed and open string variations is schematically 
\def\dcl{\de_z}\def\dop{\de_\zh}
\eqn\ocdia{
\xymatrix{
 F^3 H^3(Z^*)\ar[r]^\dcl\ar[dr]^\dop 
&F^2 H^3(Z^*) \ar[r]^\dcl\ar[dr]^\dop 
&F^1 H^3(Z^*) \ar[r]^\dcl\ar[dr]^\dop  
&F^0 H^3(Z^*) \ar[d]^\dop 
\cr
&F^2 H^2(\cx H) \ar[r]^{\dcl,\dop}
&F^1 H^2(\cx H) \ar[r]^{\dcl,\dop} 
&F^0 H^2(\cx H) \cr 
}
}
Here $F^k$ is the Hodge filtration and $\dcl$ and $\dop$ denote the closed and open string variations, respectively. 
For more details on this structure we refer to refs.\refs{\LMW,\Jo} (see 
also \refs{\Forbes,\GHKK}). 
The variations $\de$ can be identified 
with the flat Gauss-Manin connection $\gm$, 
which captures the variation of mixed Hodge structure on the bundle with fibers the 
relative cohomology groups. The mathematical background is described in
refs. \refs{\mhvi,\mhvii,\mhviii}.

The flatness of the Gauss-Manin connection leads to a non-trivial 
''$\cx N=1$ special geometry''
of the combined open/closed field space, that governs the open/closed chiral ring of the 
topological string theory \LMW. This 
geometric structure leads to a Picard-Fuchs
system of differential equations satisfied by the relative period integrals 
\eqn\epf{
\cx L_a \Pi_\Sigma = 0,\qquad \Pi_\Si(z,\zh)=\int_{\ga_\Si} \Om,\qquad \ga_\Si\in H_3(Z^*,\cx H) \ .
}
Here $\{\cx L_a\}$ is a system of linear differential operators, 
$z$($\zh$) stands collectively for the closed (open) string parameters and 
the holomorphic 3-form $\Om$ and its period integrals are
defined in relative cohomology. The relative periods $\Pi_\Si(z,\zh)$ determine the mirror
map and the combined open/closed string superpotential, which can be written in a
unified way as
\eqn\supi{
\cW_{\cx N=1}(z,\zh)=\cW_{closed}(z)+\cW_{open}(z,\zh)=
\sum_{\ga_\Si \in H^3(Z^*,\cx H)}N_\Si\Pi_\Si(z,\zh)\ .
}
Here $\cW_{closed}(z)$ is the closed string superpotential proportional to the
periods over cycles $\ga_\Si\in H^3(Z^*)$ and $\cW_{open}(z,\zh)$ the brane superpotential
proportional to periods over chains $\ga_\Si$ with non-empty boundary $\p \ga_\Si$. 
The coefficients $N_\Si$ are
the corresponding ``flux'' and brane numbers.\foot{To 
obtain the physical superpotential, an appropriate choice of 
reference brane has to be made for the chain integrals, since a  
relative period more precisely computes the brane tension of a domain 
wall \refs{\WitCh,\Katz,\AV}. 
This should be kept in mind in the 
following discussion where we simply refer to ``the superpotential''.}

In the following we implement this general structure
for the class of toric branes on compact manifolds defined in sect.~2.1. 
In the present context, the deformations of $C$ are controlled by eq.\lagb\
and the relative cohomology problem is naturally defined by the hypersurfaces $\cx B_a(E)$
in the $B$ model.
In \LMW\ this identification was used to set up
the appropriate problem of mixed Hodge structure for branes 
in non-compact CY manifolds and to compute the Picard-Fuchs system
of the $\cx N=1$ special geometry. This approach was extended to the
compact case in \Jo\ by relating $\cx H$ to the 
algebraic Chern class $c_2(E)$ of a $B$ brane as obtained from
a matrix factorization. As observed in sect.~2,
these two definitions of $\cx H$ are closely related
and it is straightforward to check that they coincide in concrete examples;
in particular the hypersurfaces defined in \Jo\ fit into the definition 
of $\cx H$ via \lagb\ in \LMW.\foot{As
was stressed in sect.~3.6 of \MW, the 
chain integrals, which define the {\it normal functions} associated with 
the superpotential, do not depend 
on the details of the infinite complexes constructed in \Heho.
Our results suggest that the relevant information for the superpotential 
is 
captured by the linear sigma model defined below.}

\subsec{GLSM and enhanced toric polyhedra}
To make full use of the machinery of toric geometry we start with 
defining a GLSM for the CY/brane geometry. 
The GLSM puts the CY geometry and the brane geometry
on equal footing and allows to study the phases of the combined system 
by standard methods of toric geometry .
The GLSM thus provides valuable
information on the {\it global} structure of the combined open/closed 
deformation space which will be important 
for identifying and investigating the various phases of 
the brane geometry, in particular large volume phases. 

We will use the concept of toric polyhedra to define the GLSM for the 
mirror pairs of toric brane geometries. This approach has the advantage of
giving a canonical construction of the $B$ model mirror to a certain
$A$ brane geometry and provides a short-cut to derive the 
generalized hypergeometric system for the relative periods given in eq.$(2.18)\yyy$ below.
As discussed above, Batyrev's correspondence describes a mirror pair of
toric hypersurfaces $(Z,Z^*)$ by a pair of dual polyhedra $(\D,\Ds)$. 
What we are proposing here is that there is a similar correspondence between 
``enhanced polyhedra'' $(\De(Z,L),\Des(Z^*,E))$ and the pair $(Z,Z^*)$ of mirror
manifolds 
{\it together} 
with the pair of mirror branes $(L,E)$ as defined before. 

The enhanced polyhedron $\De(Z,L)$ has the
following simple structure: The points $\nu_i(Z)$ of $\D(Z)$ defining the manifold $Z$ 
are a subset of the points of $\De(Z,L)$ that lie on a hypersurface $H$ in a five-dimensional
lattice $\M_5$. We choose an ordering of the points $\mu_i\in \De(Z,L)$ 
and coordinates on $\M_5$ such that the points in $H$ are given by
$$
(\mu_i)=(\nu_i,0),\ i=1,...,k\ ,
$$
where $k$ is the number of points of $\D(Z)$.
The brane geometry is described by $k'$ extra points $\nuh_i$ with $(\nuh_i)_5<0$, where
$k'$ is related to the number $\hx n$ of (obstructed) moduli of the brane by
$k'=\hx n+1$. Thus $\De(Z,L)$ is defined as the convex hull of the points 
\def\conv{{\rm conv\, }}

\eqn\defDe{
\De(Z,L) = \conv\big(\{\mu_i(\D(Z))\} \cup \{\nuh_i(L)\}\big),
\qquad \{\mu_i(\D(Z))\}\subset \De(Z,L) \cap H \ ,
}

\noi For simplicity we assume that the polyhedron $\Des$ can be 
naively defined as the dual of $\De$ in the sense of \Bat. 

To make contact between the definition of the toric branes in sect.~2
and the extra points $\nuh_i$, consider the 
linear dependences between the points of $\De(Z,L)$ 
\eqn\relsde{
\sum_i \lb^a_i(\De)\mu_i=0\ .
}
These relations may be split into two sets in an obvious way. 
There are $h^{1,1}(Z)$ relations, say 
$$(\lb^a(\De))=(l^a(\D),0^{k'}),\ a=1,...,h^{1,1}(Z)\ ,$$
which involve only the first $k$ points and reflect the original relations 
$l^a(\D)$ between the points $\nu_i(Z)$ of $\D(Z)$; 
they correspond to K\"ahler classes of the manifold $Z$.
The remaining relations $\lb^a(\De)$, $a>h^{1,1}(Z)$ 
involve also the extra points $\nuh_i$. To describe a brane as defined by 
the charge vectors $\lh^a(L)$ we choose the points $\nuh_i$
such that the remaining relations are of the form
$$
(\lb^a(\De))=(\lh^{a}(L),...),\ a>h^{1,1}(Z)\ .
$$
The above prescription for the construction of the enhanced polyhedron 
$\De(Z,L)$ from the polyhedron
$\D(Z)$ for a given manifold $Z$ and the definition \laga\ of the $A$ brane $L$ in sect.~2 is
well-defined if we require a minimal extension by $k'=\hx n+1$ points.

\subsec{Differential equations on the moduli space}
The combined open/closed string deformation space of the brane geometries
$(Z,L)$ or $(Z^*,E)$ can now be studied by standard methods of toric geometry. 
Let\foot{The
underscore on $\lb^a(\De)$ will be dropped again to simplify notation.} $\{l^a_i\}$ denote
a {\it specific} choice of basis for the generators of the relations \relsde\
in the GLSM and 
$a_i$ the coefficients of the hypersurface equation 
$p=\sum_i a_iy_i$ of the mirror $B$ model. From the homogeneous coordinates 
$a_i$ on the complex moduli space one may define local coordinates
associated with the choice of a basis $\{l^a_i\}$ by\foot{The 
sign is a priori convention but receives a meaning 
if the classical limit of the mirror map is fixed as in \hkty.}
\eqn\defmod{
z_a=(-)^{l^a_0} \prod_i a_i^{l^a_i}\ ,\qquad a=1,...,M+N.
}
Our main tool will be a system of linear differential equations of the form
\eqn\defpicf{
\cx L_a \Pi(z_b)=0,}
whose solutions are the relative periods \epf. 
The relative periods determine not only the genus zero partition functions but also 
the mirror map $z_a(t_b)$ between the flat coordinates 
$t_a$ and the algebraic moduli $z_a$ for the open/closed string deformation 
space \LMW.
There are two ways to derive the system of 
differential operators $\{\cx L_a\}$: 
Either as the canonical generalized hypergeometric GKZ 
system associated with the   
enhanced polyhedron $\De(Z,L)$ \refs{\rgkz, \Bat}. 
Or as the system of differential equations capturing the
variation of mixed Hodge structure
on the relative cohomology group $H^3(Z^*,\cx H)$ as
in refs. \refs{\LMW,\Jo}. 

Here we use the short-cut of toric polyhedra and define the  
Picard-Fuchs system as the canonical GKZ system associated
with $\De$.\foot{We are 
tacitly assuming that the GKZ system $\{\cx L_a\}$ is already 
a complete Picard-Fuchs system,
which is possibly only true after a slight modification of the 
GKZ system.}  
The derivation of the Picard-Fuchs system from 
the variation of mixed Hodge structure on the relative
cohomology group, which is similar to that in \LMW, will be given in \wip;
the coincidence of the two definitions is non-trivial and reflects
a string duality \refs{\PM,\wip}.
By the results of \refs{\rgkz,\Bat}, the generalized 
hypergeometric system associated to $(\De,\Des)$ 
leads to the following differential operators for $a=1,...,M+N$:

\eqn\gkz{
\cx L_a=
\prod_{k=1}^{l_0^a}(\th_{a_0}-k)\prod_{l_i^a>0}\prod_{k=0}^{l_i^a-1}(\th_{a_i}-k)
-(-1)^{l_0^a}z_a \prod_{k=1}^{-l_0^a}(\th_{a_0}-k)\prod_{l_i^a<0}\prod_{k=0}^{-l_i^a-1}(\th_{a_i}-k)
}\vskip10pt

\noi Here $\th_x$ denotes a logarithmic derivative $\th_x=x\fc{\p}{\p x}$ and 
the derivatives of the homogeneous coordinates $a_i$ on the complex 
structure moduli and the local 
coordinates \defmod\ are related by $\th_{a_i}=\sum_a l_i^a \th_{z_a}$.  
The products are defined to run over non-negative $k$ only so that the 
derivatives $\th_{a_0}$ appear only in one of the two terms 
for given $a$. 
The solutions of the Picard-Fuchs system in eq.\gkz\ have a nice
expansion around $z_a=0$; 
expansions around other points in the moduli space can be obtained
from a change of variables. 

Eqs. \defpicf,\gkz\ represent the homogeneous 
Picard-Fuchs system for the brane geometry $(Z^*,E)$.
These homogeneous Picard-Fuchs equations 
give rise to inhomogeneous Picard-Fuchs equations by
splitting the operators $\cx L_a$ in
a piece $\cx L_{a,bulk}$ that depends only on the moduli $z$  of the manifold $Z^*$ and essentially represent the Picard-Fuchs system of the CY geometry 
and a part 
$\cx L_{a, open}$ that governs the dependence on the 
open string deformations $\zh$. 
Upon evaluation at a critical point w.r.t. the open string deformations, 
$\de_{\zh}\cW=0$, the 
split leads to an inhomogeneous term $f_a(z)$, if 
$\Pi$ is a chain that depends non-trivially on the brane deformations
$\zh$. 
\vskip5pt
\eqn\Lsplit{
\cx L_{a,bulk} \Pi(z,\zh) = -\cx L_{a,open} \Pi(z,\zh) \quad
\buildrel\de_{\zh}\cW=0\over\longrightarrow\quad 
\cx L_{a,bulk} \Pi(z) =f_a(z)\, .
}\vskip5pt
\noi 
For the case of the quintic, the inhomogeneous term $f_a(z)$ has been
computed by a careful application of the 
Dwork-Griffiths reduction method for the chain integrals in \MW\
and it is straightforward to check that this term agrees with the
inhomogeneous term on the r.h.s of \Lsplit, see eq.$(3.11)\yyy$ below.

In \GHKK\ it has been proposed that the problem of mixed hodge variations 
on the relative cohomology groups defined in \refs{\LMW,\Jo} 
can be reinterpreted in terms of the deformations of 
a certain non Ricci-flat K\"ahler blow up $\tx Y$ of the $B$ model geometry.
It has been further suggested that it should be possible to obtain
a Picard-Fuchs system for the brane geometry by computing 
in the manifold  $\tx Y$ and restricting the complex structure of $\tx Y$
in an appropriate way. At the moment the details 
appear to be unknown and it would be interesting to relate these ideas
to the above results. It would also be interesting to understand a
possible connection to the differential equations and superpotentials
derived from matrix factorizations in \refs{\BBG,\KSi,\KSii}.

\subsec{Phases of the GLSM and structure of the solutions of \gkz}
In the previous definitions we have used a specific choice of basis $\{l_i^a\}$ 
to define the local coordinates \defmod\ and the differential operators \gkz.
Different choices of coordinates correspond to different phases of the GLSM 
\glsm. The extreme cases are on the one hand 
a large volume phase in all the K\"ahler parameters, 
where the GLSM describes a smooth classical geometry
and on the other hand a pure Landau-Ginzburg phase. In between
there are mixed phases, where only some of the moduli are at large volume and
other moduli are fixed in a stringy regime of small volume. A nice instanton expansion
can be expected a priori only for moduli at large volume. 

Representing the GLSM by the toric polyhedron $\De$, the
different phases of the GLSM may be studied by considering different
triangulations of the polyhedron \refs{\glsmii,\Bat}.
Without going into the technical details of this procedure, 
let us outline the relevance of this phase 
structure in the present context. A given $B$ brane configuration corresponds to a 
critical point of the superpotential which lies 
in a certain local patch of the parameter space. 
To study the 
critical points in a given patch and to give a nice local expansion of the 
superpotential it is necessary to work in the appropriate local 
coordinates. The different triangulations of $\De$ 
define different regimes in the parameter space, where the relative
periods $\Pi_\Si$ have a certain characteristic behavior depending 
on whether the brane moduli are at large or at small volume. 
To find an interesting instanton expansion we
look for triangulations that correspond to patches where at least 
some of the moduli are at large volume.

From the interpretation of the system $\{\cx L_a\}$ of differential operators 
as the Picard-Fuchs system for the relative periods on $Z^*$ we expect the
solutions of the equations \defpicf\ to have the following structure:

\item{$a)$} There are $2M+2$ solutions $\Pi(z)$ that represent the periods of $Z^*$
up to linear combination and depend only on the complex structure 
moduli $z_a$, $a=1,...,h^{1,1}(Z)$ of $Z^*$.

\item{$b)$} There are $2N$ further solutions $\hx \Pi(z,\hx z)$ that do depend on all deformations
and define the mirror map for 
the open string deformations and the superpotential (more precisely: brane tensions).

\item{$c)$}
\hskip-6pt\foot{The following holds
for appropriate choices of normalization 
and the sign in \defmod\ that have been made in 
\gkz, explaining the special appearance of the entry $i=0$ corresponding
to the fiber of the anti-canonical bundle.}%
For a maximal triangulation corresponding to a large complex structure point 
centered at 
$z_a=0\ \forall a$, there will be a 
series solution $\om_0(z_a)=1+\cx O(z_a)$ and $M+N$ solutions 
$\om_c(z_a)$ with a single log behavior
that define the open/closed mirror maps as ($c$ is fixed in the following equation)
$$
t_c(z_a)=\fc{\om_c(z_a)}{\om_0(z_a)}=\fc{1}{2\pi i }\ln(z_c)+S_c(z_a),
$$
where $S_c(z_a)$ is a series in the coordinates $z_a$.

\noi It follows from $a)$ that the mirror map $t\cl(z)$ in the closed string
sector does not involve the open string deformations, similarly as has been
observed in \refs{\AV,\AKV,\LMW} in the non-compact case.\foot{This statement
holds at zero string coupling.}
However the open string 
mirror map $t\op(z,\zh)$ depends on both types of moduli. For 
explicit computations of the mirror maps at various points in the
moduli we refer to the examples.

The special solution $\Pi=\cW_{open}(z,\zh)$ has the further property that 
its instanton expansion near a large volume/large complex structure point
encodes the Ooguri-Vafa invariants of the brane geometry:
\eqn\wov{
\Winst(q_a)=\sum_\be G_\be \vq^\be = \sum_\be \sum_{k=1}^\infty 
N_{\vbe} \fc{\vq ^{k.\be}}{k^2}\ .
}
Here $\vbe$ is the non-trivial homology class of a disc, $\be \in H^2(Z,L)$, $\vq^\be$
a weight factor related to its appropriately defined 
K\"ahler volume, $G_\be$ the fractional Gromov-Witten type coefficients in the 
instanton expansion and $N_\be$ the integral Ooguri-Vafa invariants \OV. 

Below we study some illustrative examples and
find agreement with the above expectations.

\newsec{Applications}
In the following we apply the above method to study some 
examples including involution branes with obstructed 
deformations as well as a class of branes with classically unobstructed 
moduli. 
 
\subsec{Branes on the quintic $\IX^{(1,1,1,1,1)}_5$}
\def\zcl{z_1^{(cl)}}
\def\xcl{x_1^{(cl)}}
\def\Wc{\cW_{crit}}
\vskip-18pt
\subsubsec{\mn.1. Brane geometry}
We first study a family of toric branes on the quintic that 
includes branes that have been studied before in \refs{\Wa,\MW,\Jo}
by different means. We recover these results for special choice of 
boundary conditions and study connected configurations.
As in sect.~2. we consider 
a one parameter family of $A$ branes 
defined by the two charge vectors
\eqn\exi{
(l^1)=(-5,1,1,1,1,1),\qquad (\lh^2)=(1,-1,0,0,0,0)\ .
}
As discussed in sect.~2.3 we may associate with this brane
geometry a five-dimensional toric polyhedron $\De(Z,L)$ that contains 
the points of the polyhedron $\D(Z)$ of the quintic as a subset 
on the hypersurface $y_5=0$:
\vs

\vbox{
$$\vbox{\offinterlineskip\halign{
# height 15pt depth 4pt&~$#$\hfil\vrule
&~$#$\hfil&\hfil\quad$#$\quad\hfil\vrule\cr
\noalign{\hrule}
\vrule&\D(Z) 
&	\nu_0=& (\- 0,\- 0,\- 0,\- 0,\- 0)	\cr
\vrule&&\nu_1=& ( - 1,\- 0,\- 0,\- 0,\- 0)	\cr
\vrule&&\nu_2=& (\- 0,  -1,\- 0,\- 0,\- 0)	\cr
\vrule&&\nu_3=& (\- 0,\- 0,  -1,\- 0,\- 0)	\cr
\vrule&&\nu_4=& (\- 0,\- 0,\- 0,  -1,\- 0)	\cr
\vrule&&\nu_5=& (\- 1,\- 1,\- 1,\- 1,\- 0)	\cr
\noalign{\hrule\vskip3pt\hrule}
\vrule&\hfil \De(Z,L)=\D\cup\ 
	&\rho_1=& (  -1,\- 0,\- 0,\- 0, -1)	\cr
\vrule& &\rho_2=& (\- 0,\- 0,\- 0,\- 0, -1)	\cr
\noalign{\hrule}
}}
$$
\nobreak\tablecaption{1}{Points of the enhanced polyhedron $\De$ for the
geometry \exi\ on $\IX^{(1,1,1,1,1)}_5$.}}
\vskip10pt
\noi Choosing a maximal triangulation of $\De(Z,L)$ determines the following
basis of generators for the relations \relsde\foot{The following 
computations have been performed using parts of existing computer codes \comp.}
\eqn\eximori{
l^1=(-4,0,1,1,1,1;1,-1),\qquad l^2=(-1,1,0,0,0,0;-1,1)\ ,
}
where the last two entries correspond to the extra points. In the
local variables\foot{We equipped $z_1$ with an additional minus sign compared 
to \defmod\ for later convenience.}
\eqn\eximod{
z_1=-\fc{a_2a_3a_4a_5a_6}{a_0^4a_7},\qquad z_2=-\fc{a_1a_7}{a_0a_6},
}
the hypersurface equations for the $B$ brane geometry $(Z^*,E)$ read
\def\zcl{z}\eqn\exibm{
\eqalign{
p(Z^*)&:\ x_1^5+x_2^5+x_3^5+x_4^5+x_5^5-x_1x_2x_3x_4x_5\, \zcl^{-\fc{1}{5}} =0,\cr
\cx B(E)&:\ x_1^5+x_1x_2x_3x_4x_5\ z_2\zcl^{-\fc{1}{5}}=0 \ . 
}}
Here $\zcl=-z_1z_2$ denotes the complex structure modulus of the
CY geometry $Z^*$. 

From eq.\eximori\ one can immediately proceed and solve 
the toric Picard-Fuchs system \gkz\ to derive
the mirror maps and the superpotentials and we will do so momentarily. 
However it is instructive to 
take a closer look at the geometry of the problem of mixed Hodge variations on 
the relative cohomology groups \ocdia, which has the following 
intriguing structure. Rewriting the superpotential $p(Z^*)$ in the
original variables $y_i$ of the toric ambient space and restricting
to the hypersurface $\cx B(E):\ y_1=y_0$ in these variables (cpw. \lagb)
defines the following boundary superpotential $\Wh=p(Z^*)|_{y_1=y_0}$ 
for the relative cohomology problem on $\cx H=\cx B(E)$:
$$
\Wh=(a_0+a_1)y_0+a_2y_2+a_3y_3+a_4y_4+a_5y_5 \ .
$$
The boundary superpotential $\Wh$ describes 
a K3 surface defined as a quartic polynomial in $\IP^3$ after the 
transformation of variables $y_i=x_i^4$, $i=1,...,4$:
\eqn\defkt{
\Wh=x_1^4+x_2^4+x_3^4+x_4^4+\zk^{-1/4}\ x_1x_2x_3x_4\ .
}
Thus the part of the Hodge variation associated with the lower row in \ocdia,
which can be properly defined as a subspace through the weight 
filtration \refs{\LMW,\Jo}, is the usual 
Hodge variation associated with the complex structure 
of the family of K3 manifolds defined by $\Wh$.
The complex structure determined by the (2,0) form $\om$
on the K3 is parametrized by the modulus 
\def\larrow{\ -\hskip-5pt-\hskip-5pt-\hskip-5pt-\hskip-5pt\longrightarrow\ }
$$
\zk=\fc{z_1}{(1+z_2)^4}\buildrel a_6/a_7=-1\over{\larrow} \fc{a_2a_3a_4a_5}{(a_0+a_1)^4}  \ ,
$$
which is a special combination of the closed and open string moduli.
Since the dependence of the Hodge variation on the brane 
modulus $z_2$ localizes on $\cx H$, the open
string mirror map and the brane tension will be directly related 
to periods on the K3 surface \defkt! This observation is very useful
in studying details of the critical points and generalizes to 
other brane geometries \wip.

\def\si{\sigma}\def\Lxt{\tilde{\cx L}}
The differential operators \gkz\ in the local variables $z_1,z_2$ 
defined by \eximori\ read 
\eqn\exila{\eqalign{
\cx L_1 &= (\th_1^4-z_1 \prod_{i=1}^4(4\th_1+\th_2+i)) (\th_1-\th_2)\ ,\cr
\cx L_2 &= (\th_2+z_2(4\th_1+\th_2+1))(\th_1-\th_2)\ .}}
The above operators $\cx L_1$ and $\cx L_2$ reveal the relation of the 
variation of mixed Hodge structure 
to the family of K3 manifolds defined in \defkt. 
Indeed the combination $(\th_1-\th_2)$ is the direction of the open 
string parameter that localizes on $\cx H$. The split 
$$\cx L_a = \Lxt_a (\theta_1-\theta_2)\, , $$ shows that 
the solutions $\pi_\si$ of the equations $\Lxt_a\pi_\si=0$ are just the K3 periods.
The operator $\Lxt_2$ imposes that the periods depend non-trivially only
on the variable $\zk\!$\foot{The $z_2$ dependent prefactor arises from the normalization 
of the holomorphic form.}
$$ \Lxt_2 \left((z_2+1)^{-1} f(\zk) \right)=0 \, ,$$ 
whereas the operator $\Lxt_1$ reduces to the 
Picard-Fuchs operator of the K3 surface in the new variable $\zk$. 
 It follows that the solutions 
of the K3 system are the first variations of the relative periods w.r.t.
the open string deformation  and 
a critical point $\hx \delta W=0$ corresponds to a particular solution $\pi$ 
of the K3 system that vanishes at that point. 
The solution that describes the involution brane is determined by requiring the
right transformation property under the discrete symmetry of the moduli
space as in \Jo. 

Further differential operators 
can be obtained from linear combinations of the basis vectors $l^a$. E.g. the linear
combination $l=l^1+l^2$ defines the differential operator
$$
\cx L_1' =  \th_2  \th_1^4  + z_1 z_2 \prod_{i=1}^5 (4\th_1+\th_2+i) \ ,
$$
which also annihilates the relative periods.\foot{One can further factorize the
above operators to a degree four differential operator 
which together with $\cx L_2$ represents a complete Picard-Fuchs system.}
The solutions of the complete system of differential operators 
have the expected structure described in sect.~2.5. The mirror maps can be computed to be
\eqn\eximm{\eqalign{
-z_1(t_1,t_2)&= \, q_1  + (24 \, q_1 ^2 - \, q_1  \, q_2 ) + 
     (-396 \, q_1 ^3 - 640 \, q_1 ^2 \, q_2 ) 
+\ldots\ ,\cr
z_2(t_1,t_2)&= q_2  + (-24 \, q_1  \, q_2  + \, q_2 ^2) + 
     (972 \, q_1 ^2 \, q_2  - 178 \, q_1  \, q_2 ^2 + \, q_2 ^3) 
+\ldots\ ,\cr
}
}
with $q_a=\exp (2\pi i\,  t_a)$.
The deformation parameters $t_1$ and $t_2$ are the flat coordinates near 
the large complex structure point $z_1=z_2=0$ 
associated with {\it 
open string} deformations \LMW. Their physical interpretation is the 
quantum volume of two homologically distinct discs 
as measured by the tension of D4 domain walls
on the $A$ model side \refs{\AV,\AKV}. 
The other solutions of the differential operators \gkz\ describe
the brane tensions \supi\ of the domain walls in the family.
We proceed with a study of various critical points of the superpotential.

\vskip-8pt
\subsubsec{\mn.2. Near the involution brane}
To study brane configurations mirror to the involution brane we 
consider a critical point of the type \bbq, that is a D5 brane locus
$$
x_2^5+x_3^5=0,\qquad x_4^5+x_5^5=0,\qquad x_1^5-x_1x_2x_3x_4x_5\, \zcl^{-\fc{1}{5}}=0 \ .
$$
Comparing with \exibm\ we search for a superpotential with 
critical locus near $z_2=-1$
and arbitrary $z_1$. 
Let us first look at 
the large volume phase $z_1\sim 0$ of the mirror $A$ brane,
where one expects an instanton expansion with integral coefficients.
The local variables \eximod\ are centered at $z_1=z_2=0$,
not $z_2=-1$, however. To get a nice  expansion
of the superpotential near the locus $z_2+1=0$ we change variables to 
$$
u=z_1^{-1/4}(1+z_2) ,\qquad v=z_1^{1/4}\ .
$$ 
Examining the $z_2$-dependent solution of the GKZ system in these
variables, we find the superpotential 
\eqn\exisplg{
c\ \cx W(u,v)=\frac{u^2}{8}+15 v^2+\frac{5 u^3 v}{48}-\frac{15 u v^3}{2}+\frac{u^6}{46080}+\frac{35 v^2
   u^4}{384}-\frac{15 v^4 u^2}{8}+\frac{25025 v^6}{3} + \dots
}
which has the expected critical locus $\hx \delta \cx W=0$  
at $u=0$ for all values of $v$. Here $c$ is 
a constant that can not be fixed from the consideration of the
differential equations \gkz\ alone.\foot{The precise  
linear combination of the solutions of the Picard-Fuchs system
that corresponds to a given geometric cycle can be determined by
an intersection argument and possibly analytic continuation, similarly as
in the closed string case \can. Such an argument has been
made in the present example already in \Wa, from which we will borrow 
the correct value for $c$.} At the critical locus $u=0$ the above expression 
yields the critical value $\Wc(\zcl)=\cx W(u=0,v=\zcl^{1/4})$ 
\eqn\exiwcrit{
\Wc(\zcl)=15\,{\sqrt{\zcl}}\,+ \fc{25025}{3}\,{\zcl}^{3/2}\, + \fc{52055003}{5}\,{\zcl}^{5/2}\, + \dots \, .}
Here the constant has been fixed to $c=1$ by comparing \exiwcrit\ 
with the result of \Wa\ for $\Wc(\zcl)$.

As alluded to in sect.~2.5, the differential operators \exila\ have the 
special property that the periods of $Z^*$ are amongst their solutions. 
One may check
that the open string mirror maps $\eximm$ conspire such
that the mirror map for the remaining modulus $z=-z_1z_2$ at the critical
point coincides with the closed string mirror map for the quintic.
Using the multi-cover prescription of \refs{\OV,\Wa} and expressing 
\exiwcrit\ in terms of the exponentials 
$q(z)=\exp (2\pi i\,  t(z))=z+\cx O(z^2)$  one obtains the 
integral instanton expansion of the $A$ model 
$$\eqalign{
\fc{\Wc(\zcl(q))}{\omega_0(q)}&= 15\,{\sqrt{q}}\, + \fc{2300}{3}\,{q}^{3/2}\, + \fc{2720628}{5}\,{q}^{5/2}+ \dots  \, ,\cr
&= \sum_{k \, odd}( \fc{15}{k^2}  q^{k/2}+ \fc{765}{k^2} q^{3k/2}+ \fc{544125}{k^2} q^{5k/2} + \dots )\, .}
$$
To make contact with the inhomogeneous Picard-Fuchs equation of \MW, we rewrite the differential
operators above in terms of the bulk modulus $\zcl$ and 
the open string deformation $z_2$ and split off the $z_2$ dependent
terms as in \Lsplit. In particular the operator $\cx L_1'$ leads to a non-trivial
equation of the form $\th \cx L_{bulk}\Pi=-\cx L_{open}\Pi$, where  
\eqn\exiwcritii{\eqalign{
\cx L_{bulk}&= \th^4 -5 \zcl \prod_{i=1}^4(5 \th + i)\ ,\qquad \cx L_{open}=\cx L_1'-\th \cx L_{bulk} \, , \cr
\cx L_1' &=  (\th +\th_2)\th^4-\zcl \prod_{i=1}^5 (5 \th +\th_2+i) \, ,}}
and $\th=\th_z$.
Setting $\Pi=\cx W(u,v)$ and restricting 
to the critical locus $z_2=-1$ one obtains
\eqn\redmw{
\cx L_{bulk} \Wc = \fc{15}{16}\sqrt{\zcl}\ .
}
This identifies the inhomogeneous Picard-Fuchs equation of \refs{\Wa,\MW} 
as the restriction of \gkz\ to the critical locus.

While the result \exiwcrit\ had been previously obtained in \Wa,
the above derivation gives some extra information. Since the 
definition of the toric branes holds off the involution 
locus, the superpotential $\cx W(u,v)$ describes  more generally any 
member of the family of toric $A$ branes defined by \laga, 
not just the involution brane. It describes also the deformation
of the large volume superpotential away from $z_2=-1$. It is also possible to 
describe more general configurations with 
several deformations \wip. It should also be noted
that the use of the closed string mirror map in \Wa\ was strictly 
speaking an assumption, as the closed string mirror map measures 
the quantum volume of fundamental sphere instantons, not the
quantum tension of D4 domain walls wrapping discs, which is the appropriate 
coordinate for the integral expansion of \OV.
It is neither obvious nor true in general that this D4 tension agrees with 
half the sphere volume of the fundamental string, in particular
off the involution locus. In the present case it is not hard to justify this
choice and to check it from the computation of the mirror map, but
more generally there will be corrections to the D4 quantum
volume that are not determined by the closed string mirror map, see eq.\eximm\
and the examples below.
\vs 
\noi {\it Small volume in the $A$ model: $1/z_1\sim 0$}\br
Another interesting point in the moduli space is the Landau-Ginzburg 
point of the $B$ model. This case has been studied previously in \Jo,
so we will be very brief. The only non-trivial thing left to 
check is that the 
system of differential equations obtained in \Jo\ from 
Dwork-Griffiths reduction is equivalent to the toric GKZ system \gkz\ 
transformed to the local variables near the LG point. Choosing
local variables 
$$
x_1 = \fc{a_0}{(a_2 a_3 a_4 a_5)^{1/4}}\left(\fc{-a_7}{a_6}\right)^{1/4},\qquad x_2 = \fc{a_1}{(a_2 a_3 a_4 a_5)^{1/4}}\left(\fc{-a_7}{a_6}\right)^{5/4}\ ,
$$
one obtains by a transformation of variables the differential operators 
\eqn\exilgop{\eqalign{
\cx L_1 &= (x_1^4 (\theta_1 + \theta_2)^4 -4^4 \prod_{i=1}^4 (\theta_1-i))(\theta_1+5 \theta_2)\, , \cr
\cx L_2 &= \left( x_2 (\theta_1-1)- x_1 \theta_2 \right) (\theta_1+5 \theta_2) \, ,\cr
\cx L_1' &= x_1^5 (\theta_1 + \theta_2)^4 \theta_2 -4^4 x_2 \prod_{i=1}^5 (\theta_1-i)\, ,
}}
\def\xcl{x}%
where $\th_i$ denotes the logarithmic derivatives $\th_{x_i}$. The above operators agree with eqs.(5.14)-(5.16) 
of \Jo\ up to a change of variables. The superpotential is
$$
\cx W=-\frac{x_1^2}{2}-\frac{x_2 x_1}{6} -\frac{x_1^6}{11520}-\frac{x_2 x_1^5}{3840}-\frac{x_2^2
   x_1^4}{2688}-\frac{x_2^3 x_1^3}{3456}-\frac{x_2^4 x_1^2}{8448}-\frac{x_2^5 x_1}{49920}+\dots \, ,
$$
which has its critical locus at $x_2=-x_1$, which corresponds to $u=0$ in these coordinates. In terms of the closed string variable 
$\xcl= -x_1 x_2^{-1/5}$ at the Landau Ginzburg point, the expansion at the critical locus reads 
$$
\cx W_{crit}=-\fc{\xcl^{5/2}}{3} -\fc{ \xcl^{15/2} }{135135} -\fc{\xcl^{25/2}}{1301375075} +  \dots \, ,
$$
which satisfies a similar equation as \redmw\
$$
 \cx{L}_{bulk} \cx W_{crit}=  \fc{15}{16}\,  \xcl^{5/2} \, ,
$$
where
$
 \cx{L}_{bulk} = 5^{-4}\xcl^5 \th_x^4-5 \prod_{i=1}^4 (\th_x-i) \, .
$

\def\opt{\cx O(-3)_{\IP^2}}
\vskip20pt
\subsec{Branes on  $\IX^{(1,1,1,6,9)}_{18}$}
As a second example we study branes on the two moduli CY 
$Z=\IX^{(1,1,1,6,9)}_{18}$. $Z$ is an elliptic fibration over $\IP^2$
with the elliptic fiber and the base parametrized by 
the coordinates $x_1,x_2,x_3$ and $x_4,x_5,x_6$ in \defmfs, respectively.
In the decompactification limit of large fiber, the compact CY approximates the
non-compact CY $\opt$ with coordinates $x_3,x_4,x_5,x_6$.
This limit is interesting, as it makes contact
to the previous studies of branes on $\opt$ in \refs{\AKV,\LM}.

\subsubsec{\mn.1. Brane geometry}
We consider a family of $A$ branes parametrized by the relations
\eqn\exii{
|x_4|^2-|x_3|^2=c^1,\qquad \lh=(0,0,0,-1,1,0,0) \ .
}
This defines a family of D7-branes in the mirror parametrized by one complex modulus.
To make contact with the non-compact branes
we may add a second constraint $|x_5|^2-|x_3|^2=0$ that selects
a particular solution of the Picard-Fuchs system.\foot{Since the 
constant in this equation 
must be zero to get a non-zero superpotential \AV, 
there is no new modulus.} 
The brane geometry on the $B$ model side is defined by the two equations
\eqn\exiibm{\eqalign{
p(Z^*)&=\sum a_iy_i=a_0x_1x_2x_3x_4x_5+a_1x_1^2+a_2x_2^3+ 
a_3(x_3x_4x_5)^6+a_4x_3^{18}+a_5x_4^{18}+a_6x_5^{18},\cr
\cx B(E)&:\hskip29pt  y_3=y_4\qquad {\rm or} \qquad  (x_3x_4x_5)^6=x_3^{18}\ .
}}
As in the previous case one observes that the
complex deformations  of the brane geometry are related to the 
periods of  a K3 surface defined by 
\def\xp#1{{x'_#1}}$$
\Wh=a_0\xp1\xp2\xp3\xp4+a_1\xp1^2+a_2\xp2^3+ 
(a_3+a_4)(\xp3\xp4)^6+a_5\xp3^{12}+a_6\xp4^{12}\ .
$$
The GLSM for the above brane geometry corresponds to the enhanced 
polyhedron 
\vs

\vbox{
$$\vbox{\offinterlineskip\halign{
# height 15pt depth 4pt&~$#$\hfil\vrule
&~$#$\hfil&\hfil\quad$#$\quad\hfil\vrule
\cr
\noalign{\hrule}
\vrule&\D(Z) 
&\nu_0=& (\- 0,\- 0,\- 0,\- 0,\- 0)\cr 
\vrule&&\nu_1=& (\- 0,\- 0,\- 0,  -1,\- 0)\cr
\vrule&&\nu_2=& (\- 0,\- 0,  -1,\- 0,\- 0)\cr
\vrule&&\nu_3=& (\- 0,\- 0,\- 2,\- 3,\- 0)\cr
\vrule&&\nu_4=& (  -1,\- 0,\- 2,\- 3,\- 0)\cr
\vrule&&\nu_5=& (\- 0,  -1,\- 2,\- 3,\- 0)\cr
\vrule&&\nu_6=& (\- 1,\- 1,\- 2,\- 3,\- 0)\cr
\noalign{\hrule\vskip3pt\hrule}
\vrule&\hfil \De(Z,E)=\D\cup\ 
&\nuh_1=& (\- 0,\- 0,\- 2,\- 3, -1)\cr
\vrule&&\nuh_2=& (  -1,\- 0,\- 2,\- 3, -1)\cr
\noalign{\hrule}
}}
$$
\nobreak\tablecaption{2\yyy}{Points of the enhanced polyhedron $\De$ 
for the geometry \exii\yyy\ on $\IX_{18}$.}
\vskip10pt
}
\noi Choosing a triangulation of $\De$ that represents a 
large complex structure phase yields the following basis
of the linear relations \relsde\ between the points of $\De$:
\vskip-5pt
\eqn\exiimori{\eqalign{
\l^1=(-6,3,2,1,0,0,0,0,0),\quad
&\l^2=(0,0,0,-2,0,1,1,-1,1),\cr
&\l^3=(0,0,0,-1,1,0,0,1,-1)\ .
}}
The last two charge vectors define a GLSM for the ``inner phase'' of the 
brane in the non-compact CY described in \LM. 
The differential operators \gkz\ for the relative periods are given by
\eqn\exiidop{\eqalign{
\cx L_1 &= \th_1 (\th_1-2 \th_2-\th_3)-12 z_1 (6 \th_1+5) (6 \th_1+1),\cr
\cx L_2 &= \th_2^2 (\th_2-\th_3)+z_2 (\th_1-2 \th_2-\th_3) (\th_1-2 \th_2-1-\th_3) (\th_2-\th_3),\cr
\cx L_3 &= -\th_3 (\th_2-\th_3)-z_3 (\th_1-2 \th_2-\th_3) (\th_2-\th_3).\cr
}}

\subsubsec{\mn.2. Large volume brane}
The elliptic fiber compactifies the non-compact fiber direction $x_3$
of the non-compact CY $\opt$. 
In the limit of large elliptic fiber we therefore expect to find a deformation
of the brane studied in \refs{\AKV,\LM}. Large volume corresponds to $z_a=0$
in the coordinates defined by eqs. \exiimori,\defmod.

The mirror maps and the superpotential can be computed from \gkz. 
Expressing the superpotential in the flat coordinates $t_a$ defines
the Ooguri-Vafa invariants $N_\vbe$ in \wov. The homology class $\vbe$ 
can be labelled by three integers $(k,l,m)$ that determine
the K\"ahler volume $k t_1 + l t_2 + m t_3$ of a curve in this class.
Here $t_1$ is the volume of the elliptic fiber and $t_2,t_3$ are the
(D4-)volumes of two homologically distinct discs in the brane geometry.
The K\"ahler class of the section, which measures
the volume of the fundamental sphere in $\IP^2$,  is $t_2+t_3$. 

For the discs that do not wrap the elliptic fiber we obtain the
following invariants for $\vbe=(0,l,m)$:
\vskip8pt

\vbox{
$$\hskip-1cm\vbox{\offinterlineskip\halign{

\hfil~$#$~&\hfil~$#$~&\hfil~$#$~
&\hfil~$#$~&\hfil~$#$~&\hfil~$#$~
&\hfil~$#$~&\hfil~$#$~&\hfil~$#$~
&\hfil~$#$~&\hfil~$#$~&\hfil~$#$~
&\hfil~$#$~&\hfil~$#$~&\hfil~$#$~
&\hfil~$#$~\vrule\cr
_l\setminus ^m\vr & 0&1&2&3&4&5&6\cr
\noalign{\hrule}
0 \vr   &* &  1 &  0 &  0 &  0 &  0 &  0 \cr1\vr&
        1 &  * &  -1 &  -1 &  -1 &  -1 &  -1 \cr2\vr &
        -1 &  -2 &  * &  5 &  7 &  9 &  12 \cr3\vr &
        1 &  4 &  12 &  * &  -40 &  -61 &  -93 \cr4\vr &
        -2 &  -10 &  -32 &  -104 &  * &  399 &  648 \cr5\vr &
        5 &  28 &  102 &  326 &  1085 &  * &  -4524 \cr6\vr &
        -13 &  -84 &  -344 &  -1160 &  -3708 &  -12660 &  * \cr7\vr &
        35 &  264 &  1200 &  4360 &  14274 &  45722 &  159208 \cr8\vr &
        -100 &  -858 &  -4304 &  -16854 &  -57760 &  -185988 & -598088 \cr9\vr &
        300 &  2860 &  15730 &  66222 &  239404 &  793502 &  2530946 \cr10\vr &
        -925 &  -9724 &  -58208 &  -262834 &  -1004386 &  -3460940&  -11231776\cr
\noalign{\hrule}
}}
$$
\tablecaption{3a\yyy}{Invariants $N_{0,l,m}$ for the geometry 
\exiimori.}}\vs 
\noi The above result agrees with the results of \refs{\AKV,\LM} 
for the disc invariants in the ``inner  phase'' of the non-compact 
CY $\opt$. This result can be explained heuristically as follows.
The holomorphic discs ending on the non-compact 
$A$ brane in $\opt$ lie within the zero section of $\opt$.
Similarly discs with $k=0$ in $\IX_{18}$ are holomorphic 
curves that must map to the section $x_3=0$ of the elliptic fibration.
The moduli space of maps into the sections of the
non-compact and compact manifolds, respectively,  
does not see the compactification in the fiber, 
explaining the agreement. The agreement of the two computations 
can be viewed as a statement of local mirror symmetry in the open string
setup.

\noi For world-sheets that wrap the fiber we obtain
\vs 
\vbox{
$$\hskip-1cm\vbox{\offinterlineskip\halign{

\hfil~$#$~\vrule height 10pt depth 4pt&\hfil~$#$~&\hfil~$#$~
&\hfil~$#$~&\hfil~$#$~&\hfil~$#$~
&\hfil~$#$~&\hfil~$#$~&\hfil~$#$~
&\hfil~$#$~&\hfil~$#$~&\hfil~$#$~
&\hfil~$#$~&\hfil~$#$~&\hfil~$#$~
&\hfil~$#$~\vrule\cr
_l\setminus ^m\hskip-5pt &0&1&2&3&4&5\cr
\noalign{\hrule}
0    &* &  252 &  0 &  0 &  0 &  0 \cr1 &
        -240 &  * &  300 &  300 &  300 &  300 \cr2 &
        240 &  780 &  * &  -2280 &  -3180 &  -4380 \cr3 &
        -480 &  -2040 &  -6600 &  * &  24900 &  39120 \cr4 &
        1200 &  6300 &  22080 &  74400 &  * &  -315480 \cr5 &
        -3360 &  -21000 &  -82200 &  -276360 &  -957600 & * \cr6
&10080 &  73080 &  319200 &  1134000 &  3765000 &  13300560 \cr7 &
        -31680 &  -261360 &  -1265040 &  -4818240 &  -16380840 & \
 -54173880 \cr
\noalign{\hrule}
}}$$
\nobreak\tablecaption{3b\yyy}{Invariants $N_{1,l,m}$ for the geometry 
\exiimori.}}\vs 

\vbox{
$$\hskip-1cm\vbox{\offinterlineskip\halign{

\hfil~$#$~\vrule height 10pt depth 4pt&\hfil~$#$~&\hfil~$#$~
&\hfil~$#$~&\hfil~$#$~&\hfil~$#$~
&\hfil~$#$~&\hfil~$#$~&\hfil~$#$~
&\hfil~$#$~&\hfil~$#$~&\hfil~$#$~
&\hfil~$#$~&\hfil~$#$~&\hfil~$#$~
&\hfil~$#$~\vrule\cr
_l\setminus ^m\hskip-5pt &0&1&2&3&4\cr
\noalign{\hrule}
0    &* &  5130 &  -18504 &  0 &  0 \cr1 &
        -141444 &  * &  -73170 &  -62910 &  -62910 \cr2 &
        -28200 &  -108180 &  * &  544140 &  778560 \cr3 &
        85320 &  403560 &  1557000 &  *&  -7639920 \cr4 &
        -285360 &  -1647540 &  -6485460 &  -24088680 & \
* \cr5 &
        1000440 &  6815160 &  29214540 &  106001100 &  392435460 \cr6 &
        -3606000 &  -28271880 &  -133294440 &  -505417320 & \
 -1773714840 \cr
\noalign{\hrule}
\noalign{\hrule}
}}
$$
\nobreak\tablecaption{3c\yyy}{Invariants $N_{2,l,m}$ for the geometry 
\exiimori.}}\vs 
\noi It would be interesting to confirm some of these numbers by an independent 
computation.

\subsubsec{\mn.3. Deformation of the non-compact involution brane}
In \Waii\ an involution brane in the local model $\opt$ has been studied. 
Similarly as in the previous case one expects to find a deformation
of this brane by embedding it in the compact manifold and taking 
the limit of large elliptic fiber, $z_1=0$.
In order to recover the involution brane of the local geometry 
we study the critical points near $z_3=1$ in the local coordinates
$$ \tilde{z}_1=z_1 (-z_2)^{1/2},\quad u=(-z_2)^{-1/2} (1-z_3)\,, \quad v=(-z_2)^{1/2} \,  .$$
After transforming the Picard-Fuchs system to these variables,
the solution corresponding to the superpotential has the following expansion
\eqn\supnlib{
c \cx W= -v -\frac{35 v^3}{9}+\fc{1}{2}u v^2+\frac{200}{3} \tilde{z}_1 v^2-\frac{u^2 v}{8}-12320 \tilde{z}_1^2 v-60 u \tilde{z_1} v+ \dots\ ,
}
where $c$ is a constant that will be fixed again by comparing the critical
value with the results of \Waii. In the decompactification limit $\tilde{z}_1=0$, 
the critical point of the superpotential is at $u=0$, where we obtain the
following expansion 
\eqn\wcritopt{
c \cx W|_{crit} = - \sqrt{z_2} - \fc{35}{9} z_2^{3/2} - \fc{1001}{25} z_2^{5/2} +\dots \ ,
}
The restricted superpotential satisfies the differential equation
$$
\cx L_{bulk} \cx W|_{crit}= -\fc{\sqrt{z_2}}{8 c} \, ,
$$
with $\cx L_{bulk}$ the Picard-Fuchs operator of the local geometry $\opt$. 
The above expressions at the critical point agree with the ones given 
in \Waii\ for $c=1$.

As might have been expected, the full superpotential \supnlib\ shows that 
the involution brane of the local model is non-trivially deformed in the
compact CY manifold for $z_1\neq 0$. It is not obvious that the 
modified multi-cover description of \Wa, which is adapted to real curves and differs 
from the original proposal of \OV, 
can be generalized to obtain integral invariants for the deformations of the
critical point in the $z_1$ direction. One suspects that an integral expansion 
in the sense of \Wa\ exists only 
at critical points with an extra symmetry and for deformations that respect
this symmetry. It will be interesting to study this further.

\subsec{Branes on  $\IX^{(1,1,1,3,3)}_{9}$}
As a third example we study branes on the two moduli CY 
$Z=\IX^{(1,1,1,3,3)}_{9}$. $Z$ is again an elliptic fibration over $\IP^2$ and
one can consider a similar compactification of the non-compact brane in $\opt$.
The invariants for this geometry are reported in app.~B.

Here we consider a different family of D7-branes which we expect to 
include a brane that exists at the Landau Ginzburg point of
the two moduli Calabi--Yau. The mirror $A$ brane is defined by 
\eqn\exiii{
-|x_0|^2+|x_1|^2=c^1,\qquad \lh=(-1,1,0,0,0,0,0) \ .
}
The polyhedron for the GLSM is
\vs
\vbox{
$$\vbox{\offinterlineskip\halign{
# height 15pt depth 4pt&~$#$\hfil\vrule
&~$#$\hfil&\hfil\quad$#$\quad\hfil\vrule
\cr
\noalign{\hrule}
\vrule&\D(Z) 
&\nu_0=& (\- 0,\- 0,\- 0,\- 0,\- 0)\cr 
\vrule&&\nu_1=& (\- 0,\- 0,\- 0,  -1,\- 0)\cr
\vrule&&\nu_2=& (\- 0,\- 0,  -1,\- 0,\- 0)\cr
\vrule&&\nu_3=& (\- 0,\- 0,\- 1,\- 1,\- 0)\cr
\vrule&&\nu_4=& (  -1,\- 0,\- 1,\- 1,\- 0)\cr
\vrule&&\nu_5=& (\- 0,  -1,\- 1,\- 1,\- 0)\cr
\vrule&&\nu_6=& (\- 1,\- 1,\- 1,\- 1,\- 0)\cr
\noalign{\hrule\vskip3pt\hrule}
\vrule&\hfil \De(Z,E)=\D\cup\ 
&\nuh_1=& (\- 0,\- 0,\- 0,\- 0, -1)\cr
\vrule&&\nuh_2=& (  \- 0,\- 0,\- 0,- 1, -1)\cr
\noalign{\hrule}
}}
$$
\nobreak\tablecaption{4\yyy}{Points of the enhanced polyhedron $\De$ 
for the geometry \exiii.}
\vskip10pt
}
\noi A suitable basis of relations for the charge vectors is
\vskip-5pt
\eqn\exiiimori{\eqalign{
&\l^1=(-2,0,1,1,0,0,0,-1,1),\quad
\l^2=(0,0,0,-3,1,1,1,0,0),\cr
&\l^3=(-1,1,0,0,0,0,0,1,-1)\ ,
}}
leading to the differential operators 
\eqn\exiiidop{\eqalign{
\cx L_1 &= \th_1 (\th_1-3 \th_2) (\th_1-\th_3)+z_1 (\th_1-\th_3) (2 \th_1+1+\th_3) (2 \th_1+2+\th_3),\cr
\cx L_2 &= \th_2^3-z_2 (\th_1-3 \th_2) (\th_1-3 \th_2-1) (\th_1-3 \th_2-2),\cr
\cx L_3 &= -\th_3 (\th_1-\th_3)-z_3 (\th_1-\th_3) (2 \th_1+1+\th_3)\ .
}}
The brane geometry on the $B$ model side is defined by the two equations
\eqn\exiibm{\eqalign{
p(Z^*)&=\sum a_iy_i=a_0x_1x_2x_3x_4x_5+a_1x_1^3+a_2x_2^3+ 
a_3(x_3x_4x_5)^3+a_4x_3^{9}+a_5x_4^{9}+a_6x_5^{9},\cr
\cx B(E)&:\hskip29pt  y_0=y_1\qquad {\rm or} \qquad  x_1x_2x_3x_4x_5=x_1^{3}\ .
}}
As in the previous cases, the deformations of the 
hypersurface $\cx B(E)$ are described by the periods on a 
K3 surface.

We are interested in a brane superpotential with critical point at $z_3=-1$.
Choosing the following local coordinates centered around $z_3=-1$
$$ u=(-z_1)^{-1/2} z_2^{-1/6} (z_3 +1)\,, \quad v=(-z_1)^{1/2} z_2^{1/6} \, \quad x_2= z_2^{-1/3} \, ,$$
we obtain the superpotential 
\eqn\supxnine{
c\, \cx W=- \fc{1}{2} u x_2+ \fc{1}{24} u^3 + 210 v^3 +\fc{3}{4} v x_2^2 -\fc{3}{8} u^2 v x_2 +\dots \,.}
This superpotential has a critical point at $u=0$ and $x_2=0$.
At the critical locus we have $v=z^{1/6}$, where $z$ denotes the closed string 
modulus 
$$z=-\fc{a_1^3 a_2^3 a_4 a_5 a_6}{a_0^9}\,.$$
The expansion of the superpotential at this critical locus reads 
$$c\, \cx W|_{crit}= 210 \sqrt{z}+ \fc{53117350}{3} z^{3/2} +\fc{18297568296042}{5} z^{5/2} + \fc{7182631458065952702}{7} z^{7/2}+\dots \, ,$$
%
As in the example of sect.~3.2.3\yyy\ it is an interesting question to study 
the instanton expansion of the above expressions and its possible
interpretation in terms of integral BPS invariants. We leave this for 
the future.

\newsec{Summary and Outlook}
As proposed above, the open/closed string deformation space of the toric branes
defined in \AV\ can be studied by mirror symmetry and toric geometry 
in a quite efficient way.
The toric definition of the brane geometry in sect.~2 leads to the
canonical Picard-Fuchs system \gkz, whose solutions determine the mirror maps
and the superpotential. The phase structure of the associated GLSM determines
large volume regimes, where the superpotential has an disc instanton expansion
with an interesting mathematical and physical interpretation. 

Since the toric branes cover only a subset within the category of D-branes,
e.g. matrix factorizations on the $B$ model side, it is natural to ask for
the precise relation between these two definitions. It is an interesting 
question to which extent it is 
possible to lift the machinery of toric geometry directly to the matrix 
factorization and to make contact with the works \refs{\BBG,\KSii}.
On the positive side one notices that the class of 
toric branes is already rather large  
and not too special, as can be seen from the fact that 
the above framework covers all cases where explicit results have been obtained 
so far. 

There are some other obvious questions left open by the above discussion,
such as the geometric and physical interpretation of some of 
the objects appearing in the definition of the GLSM and the mirror $B$ geometry,
e.g. the appearance of the ``enhanced polyhedra'' $\De(Z,L)$
and K3 surfaces, which beg for an explanation. A discussion of these
issues is
beyond the scope of this paper and will be given elsewhere \wip, but here
we outline some of the answers. As the reader may 
have noticed, the polyhedra $(\De(Z,L),\Des(Z^*,E))$ define Calabi-Yau fourfolds,
which are the hallmark of F-theory compactifications with the same supersymmetry.%
\foot{An M-theory interpretation of the 4-folds for local models has been given in \PM.
The third author thanks M. Aganagic and C. Vafa for pointing out a
possible F-theory interpretation.}
Another conclusive hint towards F-theory comes from the fact that we have effectively 
studied families of 7-branes on the $B$ model side by intersecting a single 
equation with the Calabi--Yau hypersurface. In fact the 
``auxiliary geometry'' defined in sect.~2.3 should be viewed 
as a physical 7-brane geometry and this interpretation suggests that the 
results of the GLSM determine also the K\"ahler metric on the open/closed 
deformation space.
\vs\vs

\noi {\bf Acknowledgements:} We are indebted to Hans Jockers
for discussions and exchange of ideas.
We would also like to thank 
Marco Baumgartl, 
Ilka Brunner,
Thomas Grimm,
Albrecht Klemm,
Johanna Knapp,
Christian R\"omelsberger
and Emanuel Scheidegger
for discussions and comments.
The work of M.A. and P.M. is supported by the program
``Origin and Structure of the Universe'' of the German Excellence Initiative.
The work of M.H. is supported by the Deutsche Forschungsgemeinschaft.

\appendix{A}{One parameter models}
In the following we discuss the toric GKZ systems associated to brane families
connected to the involution brane in one parameter compact models.\foot{See 
\KT\ for a discussion of closed string mirror symmetry in these models.} 
At the critical value of the superpotential we recover the results of \refs{\KW,\KSi}.

\subsec{Sextic $\IX_6^{(2,1,1,1,1)}$}
We consider the charge vectors 
$$
l^1=(-4,0,1,1,1,1;2,-2),\qquad l^2=(-1,1,0,0,0,0;-1,1)\ .
$$
\subsubsec{Large volume}
This region in moduli space is parameterized by local variables
$$
 z_1=\frac{a_2 a_3 a_4 a_5 a_6^2}{a_0^4 a_7^2}  \,, \quad z_2=-\frac{a_1 a_7}{a_0 a_6} \, . 
$$
We obtain the differential operators
$$
\eqalign{
\cx L_1 &= (\theta_1^4-z_1\prod_{i=1}^4(4\theta_1+\theta_2+i)) (2\theta_1-\theta_2)\, ,\cr
\cx L_2 &= (\theta_2 + z_2 (4\theta_1+\theta_2+1))(2\theta_1-\theta_2)\, ,\cr
\cx L_1' &= \theta_1^4 \prod_{i=0}^{1}(\theta_2-i) - z_1 z_2^2 \prod_{i=1}^{6}(4\theta_1+\theta_2+i) \, . 
}$$
Switching to coordinates which are centered around the critical point
$z_2=-1$ of the superpotential 
$$
u=z_1^{-1/4} (z_2+1) \, ,\quad v=z_1^{1/4} \, ,
$$
we obtain the superpotential 
\eqn\supsextici{
c \cx W(u,v) =\frac{u^2}{24}+24 v^2+\frac{u^3 v}{24}-24 u v^3+\frac{u^6}{138240}+\frac{v^2
   u^4}{24}+\frac{143360 v^6}{3}+\dots 
}
At the critical point $u=0$, we can express $v$ in terms of 
the closed string modulus $z=z_1z_2^2$ as
$$ v|_{crit}=z^{1/4} \, .$$
We find for the superpotential at the minimum 
$$
c \cx W_{crit}=24 \sqrt{z} +\frac{143360}{3} z^{3/2} +\frac{5510529024}{25} z^{5/2} +\frac{334766662483968}{245}
   z^{7/2} + \dots \, ,
$$
This expression satisfies the differential equation
$$
 \cx L_{bulk} \cx W_{crit}= \fc{3}{2c}  \sqrt{z} \,,
$$
where $\cx L_{bulk}= \theta^4 -9 z \prod_{i=1}^4 (6\theta+i)$ denotes the Picard-Fuchs operator of the sextic. The above 
agrees with the results of \KW\ for the choice of constant $c=1$.

\subsubsec{Small volume}
To study the Landau-Ginzburg phase of the B-model we change to the local coordinates
$$ x_1 = \fc{a_0}{ (a_2 a_3 a_4 a_5)^{1/4}} \left( \fc{-a_7}{a_6}\right)^{1/2}\,, \quad x_2 =\fc{a_1}{ (a_2 a_3 a_4 a_5)^{1/4}} \left( \fc{-a_7}{a_6}\right)^{3/2}\,.$$
The differential operators obtained by a transformation of variables are ($\th_i=\th_{x_i}$)
$$
\eqalign{
\cx L_1 &= (x_1^4 (\th_1+\th_2)^4 -4^4\prod_{i=1}^4 (\th_1-i)) (\th_1+3 \th_2) \, ,\cr
\cx L_2 &= (x_2(\th_1-1)-x_1 \th_2)(\th_1+3\th_2)\, , \cr
\cx L_1'&= x_1^6 (\th_1+\th_2)^4 \th_2(\th_2-1) -4^4 x_2^2\prod_{i=1}^6 (\th_1-i) \, .
}
$$
We obtain the superpotential 
\eqn\supsexticii{
 \cx W = -\frac{1}{12} x_1^2-\frac{1}{24} x_2 x_1-\frac{x_1^6}{69120}-\frac{x_2 x_1^5}{18432}-\frac{x_2^2
   x_1^4}{11520}-\frac{x_2^3 x_1^3}{13824}-\frac{x_2^4 x_1^2}{32256}-\frac{x_2^5 x_1}{184320}+ \dots
}
which has its critical value at $x_2=-x_1$.  We can express $x_1$ in terms of the closed string variable $x=-x_1x_2^{-1/3}$ of the geometry in the Landau-Ginzburg phase as
$$
x_1|_{crit}= -x^{3/2} \,
$$ 
which gives the following critical value for the superpotential
$$
\cx W_{crit}=-\frac{x^3}{24} -\frac{x^9 }{3870720}-\frac{x^{15} }{137763225600}-\frac{5 x^{21}}{16403566461714432} +\dots
$$
This expression satisfies the equation
$$
\cx L_{bulk} \cx W_{crit} = \fc{3}{2} x^3 \,,
$$
with
$
\cx L_{bulk}= 6^{-4} x^6 \th^4 -9 (\th-1)(\th-2)(\th-4)(\th-5)\,.
$

\subsec{Octic}
We consider the charge vectors
$$
l^1=(-4,0,1,1,1,1;4,-4),\qquad l^2=(-1,1,0,0,0,0;-1,1)\ .
$$
\subsubsec{Large volume}
This region in moduli space is parameterized by local variables
$$
 z_1=\frac{a_2 a_3 a_4 a_5 a_6^4}{a_0^4 a_7^4}  \, , \quad z_2=-\frac{a_1 a_7}{a_0 a_6} \, . 
$$
The differential operators are
$$
\eqalign{
\cx L_1 &= (\theta_1^4-z_1\prod_{i=1}^4(4\theta_1+\theta_2+i)) (4\theta_1-\theta_2)\, ,\cr
\cx L_2 &= (\theta_2 + z_2 (4\theta_1+\theta_2+1))(4\theta_1-\theta_2)\, ,\cr
\cx L_1' &= \theta_1^4 \prod_{i=0}^{3}(\theta_2-i) - z_1 z_2^4 \prod_{i=1}^{8}(4\theta_1+\theta_2+i) \, . 
}$$
Switching to  $u=z_1^{-1/4} (z_2+1)$ and $v=z_1^{1/4} \, ,$ we obtain
\eqn\supoctici{
 \cx W(u,v) =\frac{u^2}{16}+48 v^2+\frac{u^3 v}{12}-96 u v^3+\frac{u^6}{92160}+\frac{5 v^2 u^4}{48}+48
   v^4 u^2+\frac{1576960 v^6}{3}+ \dots 
}
At $u=0$, we can express $v$ in terms of the classical coordinate $z=z_1z_2^4$ as $ v|_{crit}=-z^{1/4} \, .$
We find for the superpotential at the minimum
$$
c \cx W_{crit}=48 \sqrt{z}+\frac{1576960}{3} z^{3/2} +\frac{339028738048}{25} z^{5/2} +\frac{23098899711393792}{49}
   z^{7/2} + \dots \, ,
$$
which satisfies the differential equation
$$
 \cx L_{bulk} \cx W_{crit}= \fc{3}{c}  \sqrt{z} \,,
$$
where $\cx L_{bulk}=\theta^4 -16 z \prod_{i=1}^4 (8\theta+2i-1)$ denotes the Picard-Fuchs operator of the octic.
Setting $c=1$ reproduces the disk invariants of \refs{\KW,\KSi}.

\subsubsec{Small volume}
We switch to local coordinates
$$ x_1 = \fc{-a_0 a_7}{a_6 (a_2 a_3 a_4 a_5)^{1/4}} \,, \quad x_2 =\fc{a_1a_7^2}{ a_6^2(a_2 a_3 a_4 a_5)^{1/4}}\,.$$
The differential operators are ($\th_i=\th_{x_i}$)
$$
\eqalign{
\cx L_1 &= (x_1^4 (\th_1+\th_2)^4 -4^4 x_2^2 \prod_{i=1}^4 (\th_1-i)) (\th_1+2 \th_2) \, ,\cr
\cx L_2 &= (x_2(\th_1-1)-x_1 \th_2)(\th_1+2\th_2)\, , \cr
\cx L_1'&= x_1^8 (\th_1+\th_2)^4 \prod_{i=0}^3 (\th_2-i) -4^4 x_2^2\prod_{i=1}^8 (\th_1-i) \, .
}
$$
We obtain the superpotential
\eqn\supocticii{
 \cx W =-\frac{1}{16} x_1^2-\frac{1}{24} x_2 x_1-\frac{x_1^6}{92160}-\frac{x_2 x_1^5}{21504}-\frac{x_2^2
   x_1^4}{12288}-\frac{x_2^3 x_1^3}{13824}-\frac{x_2^4 x_1^2}{30720}-\frac{x_2^5 x_1}{168960}+\dots
}
At the critical value $x_2=-x_1$, we have  $x_1|_{crit}= -x^{2} \,$, where $x=-x_1 x_2^{-1/2}$. 
This gives the following expansion for the superpotential
$$
\cx W_{crit}=-\frac{x^4}{48}-\frac{x^{12}}{42577920}-\frac{x^{20}}{8475718451200}-\frac{x^{28}}{1131846085858295808}+\dots
$$
which satisfies the equation
$$
\cx L_{bulk} \cx W_{crit} = 3 x^4 \,,
$$
with
$$
\cx L_{bulk}= 8^{-4} x^8 \th^4 -16 (\th-1)(\th-3)(\th-5)(\th-7)\,.
$$
These results are in agreement with \Jo,  where this phase of the moduli space has been previously studied.

\appendix{B}{Invariants for $\IX_9^{1,1,1,3,3}$}
\noi The compactification of the local brane in $\opt$ is 
described by the charge vectors 
\eqn\exivmori{
\l^1=(-3,1,1,1,0,0,0,0,0),
\l^2=(0,0,0,-2,0,1,1,-1,1),\l^3=(0,0,0,-1,1,0,0,1,-1)\ .
}
Some invariants for this geometry are \vs
\vbox{
$$\hskip-1cm\vbox{\offinterlineskip\halign{

\hfil~$#$~&\hfil~$#$~&\hfil~$#$~
&\hfil~$#$~&\hfil~$#$~&\hfil~$#$~
&\hfil~$#$~&\hfil~$#$~&\hfil~$#$~
&\hfil~$#$~&\hfil~$#$~&\hfil~$#$~
&\hfil~$#$~&\hfil~$#$~&\hfil~$#$~
&\hfil~$#$~\vrule\cr
\phantom{\vr}&&&&k=0\cr
_l\setminus ^m\vr&0&1&2&3&4&5\cr
\noalign{\hrule}
0\vr&* &  3 &  0 &  0 &  0 &  0 \cr 
1\vr&        3 &  * &  -3 &  -3 &  -3 &  -3 \cr2\vr &
        -3 &  -6 &  * &  15 &  21 &  27 \cr3\vr &
        3 &  12 &  36 &  * &  -120 &  -183 \cr4\vr &
        -6 &  -30 &  -96 &  -312 &  * &  1197 \cr5\vr &
        15 &  84 &  306 &  978 &  3255 &  * \cr6\vr &
        -39 &  -252 &  -1032 &  -3480 &  -11124 &  -37980 \cr7\vr &
        105 &  792 &  3600 &  13080 &  42822 &  137166 \cr
\noalign{\hrule}
}}
$$
}
\vbox{
$$\hskip-1cm\vbox{\offinterlineskip\halign{

\hfil~$#$~&\hfil~$#$~&\hfil~$#$~
&\hfil~$#$~&\hfil~$#$~&\hfil~$#$~
&\hfil~$#$~&\hfil~$#$~&\hfil~$#$~
&\hfil~$#$~&\hfil~$#$~&\hfil~$#$~
&\hfil~$#$~&\hfil~$#$~&\hfil~$#$~
&\hfil~$#$~\vrule\cr
\phantom{\vr}&&&k=1\cr
_l\setminus ^m\vr&0&1&2&3\cr
\noalign{\hrule}
0\vr    &* &  27 &  0 &  0 \cr1\vr &
        -72 &  * &  90 &  90 \cr2\vr &
        72 &  234 &  * &  -684 \cr3\vr &
        -144 &  -612 &  -1980 &  * \cr4\vr &
        360 &  1890 &  6624 &  22320 \cr5\vr &
        -1008 &  -6300 &  -24660 &  -82908 \cr6\vr &
        3024 &  21924 &  95760 &  340200 \cr7\vr &
        -9504 &  -78408 &  -379512 &  -1445472\cr
\noalign{\hrule}
}}
\ 
\vbox{\offinterlineskip\halign{

\hfil~$#$~&\hfil~$#$~&\hfil~$#$~
&\hfil~$#$~&\hfil~$#$~&\hfil~$#$~
&\hfil~$#$~&\hfil~$#$~&\hfil~$#$~
&\hfil~$#$~&\hfil~$#$~&\hfil~$#$~
&\hfil~$#$~&\hfil~$#$~&\hfil~$#$~
&\hfil~$#$~\vrule\cr
\phantom{\vr}&&&k=2\cr
\vr&0&1&2&3\cr
\noalign{\hrule}
 \vr   &* &  81 &  -108 &  0 \cr\vr &
        -1269 &  * &  -1539 &  -1377 \cr\vr &
        -684 &  -2808 &  * &  13554 \cr\vr &
        2268 &  11232 &  42336 &  * \cr\vr &
        -7848 &  -46656 &  -182916 &  -671922 \cr\vr &
        27972 &  194832 &  835758 &  3020382 \cr\vr &
        -102024 &  -813456 &  -3844512 &  -14554242 \cr\vr &
        377784 &  3390336 &  17598600 &  70975872\cr
\noalign{\hrule}
}}
$$
\tablecaption{5\yyy}{Invariants $N_{k,l,m}$ for the geometry 
\exivmori.}}\vs

\noi The invariants for $k=0$ are three times the invariants in Table~3a\yyy,
where the overall factor comes from the three global sections of the 
elliptic fibration $\IX_9$. It appears that the 
invariants for $k=1,\ l\neq 0$ are generally $3/10$ times
the invariants in Table~3b\yyy. 

Some invariants for the 
geometry \exiiimori\ in the large volume phase are

\vbox{
$$\hskip-1cm\vbox{\offinterlineskip\halign{

\hfil~$#$~&\hfil~$#$~&\hfil~$#$~
&\hfil~$#$~&\hfil~$#$~&\hfil~$#$~
&\hfil~$#$~&\hfil~$#$~&\hfil~$#$~
&\hfil~$#$~&\hfil~$#$~&\hfil~$#$~
&\hfil~$#$~&\hfil~$#$~&\hfil~$#$~
&\hfil~$#$~\vrule\cr
\phantom{vr}&&&l=0\cr
_k\setminus ^m\vr &0&1&2&3&4&5\cr
\noalign{\hrule}
0\vr    &* &  54 &  0 &  0 &  0 &  0 \cr1\vr &
        -36 &  * &  54 &  -18 &  0 &  0 \cr2\vr &
        18 &  -54 &  * &  36 &  0 &  0 \cr3\vr &
        0 &  0 &  -54 &  * &  54 &  0 \cr4\vr &
        0 &  0 &  0 &  -36 &  * &  54 \cr5\vr &
        0 &  0 &  0 &  18 &  -54 &  * \cr6\vr &
        0 &  0 &  0 &  0 &  0 &  -54 \cr7\vr &
        0 &  0 &  0 &  0 &  0 &  0\cr
\noalign{\hrule}
}}
$$
}
\vskip-10pt
\vbox{
$$\hskip-1cm\vbox{\offinterlineskip\halign{

\hfil~$#$~&\hfil~$#$~&\hfil~$#$~
&\hfil~$#$~&\hfil~$#$~&\hfil~$#$~
&\hfil~$#$~&\hfil~$#$~&\hfil~$#$~
&\hfil~$#$~&\hfil~$#$~&\hfil~$#$~
&\hfil~$#$~&\hfil~$#$~&\hfil~$#$~
&\hfil~$#$~\vrule\cr
\phantom{vr}&&&l=1\cr
_k\setminus ^m\vr &0&1&2&3&4\cr
\noalign{\hrule}
0\vr    &* &  0 &  0 &  0 &  0 \cr1\vr &
        72 &  * &  -108 &  36 &  0 \cr2\vr &
        -36 &  -1728 &  * &  2772 &  -1026 \cr3 \vr&
        -1224 &  17280 &  -80460 &  * &  243756 \cr4\vr &
        5508 &  -64800 &  340092 &  -1075140 &  * \cr 
\noalign{\hrule}
}}
\ 
\vbox{\offinterlineskip\halign{

\hfil~$#$~&\hfil~$#$~&\hfil~$#$~
&\hfil~$#$~&\hfil~$#$~&\hfil~$#$~
&\hfil~$#$~&\hfil~$#$~&\hfil~$#$~
&\hfil~$#$~&\hfil~$#$~&\hfil~$#$~
&\hfil~$#$~&\hfil~$#$~&\hfil~$#$~
&\hfil~$#$~\vrule\cr
\phantom{vr}&&l=2\cr
\vr &0&1&2&3\cr
\noalign{\hrule}
\vr    &* &  0 &  0 &  0 \cr \vr&
        -180 &  * &  270 &  -90 \cr \vr&
        108 &  7020 &  * &  -11160 \cr \vr&
        -108 &  -5832 &  -97686 &  * \cr \vr&
        -10944 &  133488 &  -588276 &  2643372\cr 
\noalign{\hrule}
}}
$$
\tablecaption{6\yyy}{Invariants $N_{k,l,m}$ for the geometry 
\exiiimori.} \break It would be interesting to check some of these predictions by an 
independent computation.}

\listrefs
\end